\documentclass[11pt,cite,epsfig,psfrag]{article}
\usepackage[utf8]{inputenc}
\usepackage{placeins}
\usepackage{geometry}
 \usepackage{graphicx}
 \usepackage{subfig}
  \usepackage{amsmath} 
  \graphicspath{{images_Novel@HP/}}
  \usepackage[usenames, dvipsnames]{color}
\usepackage{wrapfig}
\usepackage{boxedminipage}

\usepackage{fullpage}
\usepackage{amsmath}
\usepackage{amssymb}
\usepackage{graphicx}
\usepackage{mathrsfs}
\usepackage{wrapfig}
\usepackage{boxedminipage}
\usepackage{epsfig}
\usepackage{setspace}
\usepackage{float}
\usepackage{hyperref}
\usepackage{enumerate}
\usepackage{cite}
\usepackage[font=footnotesize,labelfont=rm]{caption}

\usepackage[numbers,sort&compress]{natbib}

\def\be{\begin{equation}}
\def\ee{\end{equation}}
\def\bea{\begin{eqnarray}}
\def\eea{\end{eqnarray}}

\numberwithin{equation}{section}

 \newcommand{\RN}[1]{%
   \textup{\uppercase\expandafter{\romannumeral#1}}%
 }
\begin{document}

\thispagestyle{empty}

\vskip 2cm

\begin{center}
{\Large \bf Novel relations in massive gravity at Hawking-Page transition}
\end{center}

\vskip .2cm

\vskip 1.2cm

\centerline{ \bf   Pavan Kumar Yerra \footnote{pk11@iitbbs.ac.in} and Chandrasekhar Bhamidipati\footnote{chandrasekhar@iitbbs.ac.in}
}

\vskip 7mm 
\begin{center}{ School of Basic Sciences\\ 
Indian Institute of Technology Bhubaneswar \\ Bhubaneswar, Odisha, 752050, India}
\end{center}

\vskip 1.2cm
\vskip 1.2cm
\centerline{\bf Abstract}
\vskip 0.5cm
\noindent
Sign and magnitude of Ruppeiner's curvature $R_N$ is an empirical indicator of the respective nature and strength of microstructures of a thermodynamic system.  For d-dimensional Schwarzschild black holes in AdS, $R_N$ at the Hawking-Page (HP) transition point is a universal constant. In this article we
study the nature of $R_N$ at HP transition in a massive gravity theory and show that its constancy is broken due to the dependence on graviton mass as well as horizon topology. Although, when the graviton mass reaches a critical value the HP transition is driven to zero temperature, and $R_N$ at this point again approaches a universal negative constant, independent of all parameters of the theory. The geometry of black holes in AdS close to the HP transition point is also studied via novel near horizon scaling limits, and reveals the emergence of fully decoupled Rindler spacetimes.

\newpage
\setcounter{footnote}{0}
\noindent

\baselineskip 15pt

\section{Introduction}

Investigation of Phase transitions in black holes has been a rich arena to explore as it has contributed immensely to our understanding of its microscopic and macroscopic properties. One of the celebrated transitions is of course the one proposed by Hawking and Page, which happens between a hot thermal gas and a stable large black hole in a Schwarzschild black hole in anti de Sitter (AdS) spacetime~\cite{Hawking:1982dh}. The Hawking-Page (HP) 
phase transition has another interpretation due to AdS / CFT duality as a confinement/deconfinement transition in the boundary conformal field theory (CFT). In the black hole chemistry program~\cite{Kastor:2009wy,Dolan:2011xt,Karch:2015rpa,Kubiznak:2016qmn}, where the cosmological constant $\Lambda$ is considered to be dynamical giving a pressure $P = -\Lambda/8\pi$, this can also be understood as a solid-liquid phase transition. For charged AdS black holes, there thus exists a critical region, where the first order phase transition ends in a second order one~\cite{Chamblin:1999tk,Caldarelli:1999xj,Kubiznak:2012wp}.  Close to this critical region,  the thermodynamic quantities of charged AdS black holes show special scaling behavior in terms of charge $q$, where Entropy $S\sim q^2$, Pressure $p\sim q^{-2}$, and Temperature $T\sim q^{-1}$~\cite{Kubiznak:2012wp}.  In a double scaling limit, where one approaches the near horizon region together with the large charge limit, the geometry turns out to be a fully decoupled d-dimensional Rindler space-time~\cite{Johnson:2017asf}. The emergence of decoupled space-times in the near horizon limit has given enormous insights in to the physics of branes and extremal black holes in general, through AdS/CFT dualities and might teach us novel issues from the CFT point of view. It is interesting to ask whether such a decoupled Rindler spacetime is special to the critical region or can also emerge around a first order phase transition point as well and we explore it in this work.  \\

\noindent
There is also an alternative approach, namely using Ruppeiner's thermodynamic geometry~\cite{Ruppeiner:1995zz}, which is  a powerful diagnostic tool to know and classify the nature of competing microstructure interactions, which lead to phase transitions.  The key quantity in this approach is the thermodynamic curvature, whose divergences typically reveal the points where the specific heat diverges, signally the presence of a critical point or phase transition. This method has been highly successful in understanding physics around critical points~\cite{Cai:1998ep,Quevedo:2007mj,Sahay:2010tx,HosseiniMansoori:2019jcs,Chabab:2019mlu,Banerjee:2010da,Bhattacharya:2017hfj}, especially in the context of extended thermodynamics. In this context, the analysis of thermodynamic curvature denoted as $R_N$, has revealed interesting phase structure of charged black holes in AdS, where some features of microstructures were expected to be similar to the van der Waals system, but do not hold~\cite{Wei:2019uqg,Xu:2019gqm,Ghosh:2019pwy,NaveenaKumara:2020biu,Yerra:2020oph,Dehyadegari:2020ebz,Dehyadegari:2021ieh,DasBairagya:2020rqy}.  For instance, analysis of $R_N$  shows that,
while an attractive microstructure interaction is dominant for most of the parameter regime, like the van der Waals system, there is also a contrasting feature, namely, the presence of a weak repulsive interaction for small black holes at high temperature. There is a feature of thermodynamic curvature that has not been studied enough, that is its behavior close to a first order phase transition point as a naive application of the method does not show any divergence~\cite{DasBairagya:2020rqy}. However, in the case of a d-dimensional Schwarzschild black hole in AdS, it has recently been pointed out that $R_N$  at the Hawking-Page transition point turns out to be a universal constant, depending only on the dimension of spacetime~\cite{Wei:2020kra}. The universal constant thus defines a threshold, beyond which the repulsive interactions of radiation become attractive in nature aiding the formation of black holes\cite{Wei:2020kra}. An analogous calculation for the black holes in AdS with hyperbolic topology shows that $R_N$ calculated for special zero mass solutions~\cite{Emparan:1999gf,Pedraza:2018eey,Johnson:2018amj} turns out to be a universal constant along a renormalization group flow in dual conformal field theories~\cite{Yerra:2020tzg}. \\

\noindent
The aforementioned developments motivate us to give a closer look at the HP transition point, in both d-dimensional Schwarzschild as well as corresponding black holes in general theories of gravity. The HP transition though is known to occur only when the horizon is spherical and  thermodynamic geometry was reported in~\cite{Wei:2020kra}, limited to this case. In looking for existence of HP transition for arbitrary horizon topologies, an important class of examples is black holes in massive gravity theories. One of the key reasons for choosing this system, has to do with a remarkable feature which does not have a counterpart in even general Lovelock theories, i.e., the existence of HP transition as well as critical points in black hole solutions with Ricci flat or hyperbolic horizons, apart from the spherical horizon. Such black holes in AdS with non-trivial topology are called topological black holes.  Though, thermodynamic geometry has been applied to this system before to study criticality~\cite{Chabab:2019mlu,Yerra:2020oph,Wu:2020fij}, physics close to the HP transition point has not been explored yet. Second, we show that the decoupled Rindler space-time geometry found in~\cite{Johnson:2017asf} for charged AdS black holes close to the critical region, is also present for Schwarzschild black holes as well as the corresponding solutions in massive gravity in AdS at the HP transition point. This limit is facilitated by the fact that thermodynamic quantities exhibit a scaling with the cosmological constant at the HP transition point (slightly different than the scaling at the critical point~\cite{Johnson:2017asf}). Brief motivations and developments in massive gravity theories are presented below and the summary of thermodynamic geometry approach in this and other systems is discussed in section-(\ref{3}).\\

\noindent
Einstein's General theory of relativity has been highly successful with experimental confirmation of several predictions, in particular, the recent 
observational evidence of LIGO collaboration~\cite{LIGO2017,deRham2014Review} on gravitational waves. On the other hand, there are also important phenomena, for instance, accelerated expansion of the universe and the longstanding cosmological constant problem, among others, which suggest extensions to go beyond Einstein's theory. In this regard, an appealing extension involves massive graviton theories, motivated by hierarchy problems and their usefulness in quantum theory of gravity~\cite{MassiveIb,MassiveIc}, which also seems to gel well with recent observations~\cite{Abbott}, putting novel lower limits on the graviton mass. Massive gravity theories have been historically important, which were explored starting from the early models proposed by Fierz and Paullo in 1939~\cite{Fierz1939}, which went through various modifications  with the advent of new ideas, such as, New massive gravites\cite{BDghost,Newmasssive,dRGTI,dRGTII}. These theories have been actively studied in current literature \cite{NewM1,NewM2,NewM3,NewM4,NewM5,HassanI,HassanII}. Novel Black hole solutions, together with study of their thermodynamical properties~\cite{BHMassiveI,BHMassiveII,BHMassiveIII,BHMassiveIV} and applications to cosmology/astrophysics situations are also being explored with interest, to identify possible deviations from Einstein's theory~\cite{Katsuragawa,Saridakis,YFCai,Leon,Hinterbichler,Fasiello,Bamba}. An important class of massive gravity theories with applications to holographic dualities in mind was analyzed in~\cite{Vegh}, which involves a singular metric and pointing out in particular that massive
gravity may be stable and also free of ghosts \cite{HZhang}, apart from the existence of black hole 
solutions\cite{PVMassI,PVMassII,PVMassIII,PVMassIV,PVMassV}. There are other advantages. For instance, massive gravity theories hold promise in efforts to solve some of the mentioned problems in Einstein's theory as noted from~\cite{Gumrukcuoglu,Gratia,Kobayash,DeffayetI,DeffayetII,DvaliI,
DvaliII,Will,Mohseni,GumrukcuogluII,NeutronMass,Ruffini}: such as, the possibility to explain 
the current observations coming from dark matter
\cite{Schmidt-May2016DarkMatter} and also relating to the accelerating
expansion of universe without the requirement of any dark energy component
\cite{MassiveCosmology2013,MassiveCosmology2015}. It should be mentioned that attempts to embed massive gravities in string theory have been explored as well~\cite{MGinString2018}. More importantly, it has been shown that van der Waals type liquid gas phase transitions and HP transition exist in the extended phase space approach as studied in a number of works in massive black hole chemistry~\cite{Cai2015,PVMassV,PVMassIV,Alberte,Zhou,Dehyadegari,Magmass,Hendi:2016vux,Hendi:2017fxp,Dehghani:2019thq}. \\

\noindent
Summary and organization of the rest of the paper is as follows. In section-(\ref{2}), we start by writing down important thermodynamic relations for black holes in AdS in massive gravity theories, which is followed by the discussion of their phase structure, including the HP transition for various topologies in subsection-(\ref{phase}). In section-(\ref{3}), we start by obtaining a general line element on thermodynamic phase space with temperature and volume as fluctuating coordinates and use this to obtain an analytic expression for the associated thermodynamic curvature $R_N$. Analysis of thermodynamic curvature at the HP transition point i.e., $R_N\big|_{HP}$,  reveals novel features, which have no counter part in the massless limit of Einstein gravity. For example,
the universal constancy of $R_N\big|_{HP}$ noted in\cite{Wei:2020kra}, is explicitly broken in massive gravity theories. Interestingly, $R_N\big|_{HP}$ turns out be a constant at a critical value of graviton massive where the HP transition happens at zero temperature. The nature, as well as the strength of interaction depends on horizon topology as well as varies drastically with massive gravity parameters at HP transition. In particular, there is a critical pressure (or cosmological constant value) depending on massive gravity parameters, which governs whether the microstructure interactions are effectively attractive, repulsive or in balance (a point where $R_N\big|_{HP}$ vanishes). 
In section-(\ref{4}), we take a novel near horizon limit at the HP transition point of Schwarzschild black holes in AdS as well as their counterparts with non-trivial topology in massive gravity theory, to show the existence of the decoupled Rindler geometry found earlier for charged systems in\cite{Johnson:2017asf}. We end with a discussion of our findings in section-(\ref{conclusions}).

\section{AdS black holes in massive gravity} \label{2}
We start with the action for a four dimensional theory of gravity with a negative cosmological constant $\Lambda$, together with mass term $m$ for the graviton~\cite{Cai2015,PVMassIV,Hendi:2017bys}:
\begin{equation}
I=-\frac{1}{16\pi }\int d^{4}x\sqrt{-g}\left( \mathcal{R}-2\Lambda
+m^{2}\sum_{i}^{4}c_{i}\mathcal{U}_{i}(g,f)\right) .
\end{equation}
Here,  $\mathcal{R}$ is the Ricci scalar and $c_{i}$'s are constants and $f$ is a reference metric.  The quantities $\mathcal{U}_{i}$'s are formed as symmetric polynomials of the eigenvalues of a $4\times 4$ matrix $ \mathcal{K}_{\nu }^{\mu }=\sqrt{g^{\mu \alpha }f_{\alpha \nu }}$, which are in turn given as:
\begin{eqnarray}
	\mathcal{U}_{1}&=&\left[ \mathcal{K}\right], \nonumber \\ \mathcal{%
		U}_{2} &= & \left[ \mathcal{K}\right] ^{2}-\left[ \mathcal{K}^{2}\right] , \nonumber \\ 
	\mathcal{U}_{3}&=&\left[ \mathcal{K}\right] ^{3}-3\left[ \mathcal{K}\right] %
	\left[ \mathcal{K}^{2}\right] +2\left[ \mathcal{K}^{3}\right], \nonumber \\ 
	\mathcal{U}_{4}&=&\left[ \mathcal{K}\right] ^{4}-6\left[ \mathcal{K}^{2}%
	\right] \left[ \mathcal{K}\right] ^{2}+8\left[ \mathcal{K}^{3}\right] \left[
	\mathcal{K}\right] +3\left[ \mathcal{K}^{2}\right] ^{2}-6\left[ \mathcal{K}%
	^{4}\right].
\end{eqnarray}
It is well known that the above system admits static topological black hole solution, whose metric is given as~\cite{Cai2015,PVMassIV,Hendi:2017bys}:
\begin{equation}
ds^{2}=-Y(r) dt^{2}+\frac{dr^{2}}{Y(r) } + r^{2}h_{ij}dx_{i}dx_{j} \ ,
  \label{eq:metric}
\end{equation}
where $f_{\mu \nu}$ appearing above stands for the reference metric:
\begin{equation} f_{\mu \nu
}=\text{diag}(0, 0, c_0^{2} h_{ij}) \ .
\end{equation}
Here, $c_0$ is a positive constant and $h_{ij}dx_{i}dx_{j}$ is a spatial metric with
constant curvature $2k$ and volume $4\pi$, where $i,j=1,2$. The constant $k$ can take different values, such as,  $+1, 0$, or $-1$, in which cases, we obtain respectively black holes with horizons of topology, spherical, Ricci flat, or hyperbolic. Considering a particular form of reference metric $f_{\mu \nu}$ prescribed in~\cite{Cai2015,PVMassIV} , the $\mathcal{U}_{i}$'s are simplified to:
\begin{equation}
\mathcal{U}_{1}= \frac{2c_0}{r}, \quad \mathcal{%
	U}_{2} =  \frac{2c_0^2}{r^2} , \quad
\mathcal{U}_{3}= 0 , \quad
\mathcal{U}_{4}= 0 \ ,
\end{equation}
where now one can set  $c_3 =c_4 = 0$, as $\mathcal{U}_{3}= \mathcal{U}_{4}= 0$. The lapse function $Y(r)$ is~\cite{Cai2015,PVMassIV,Hendi:2017bys}:
\begin{equation}
Y(r) =k-\frac{m_{0}}{r}-\frac{\Lambda r^{2}}{3}+m^{2}(\frac{c_0c_{1}}{2}r+c_0^{2}c_{2}) \ ,
 \label{Y(r)}
\end{equation}
with the integration constant  $m_0$  corresponding to the mass $M$  of the black hole.
The solution in eqn. \eqref{Y(r)}, is asymptotically AdS and if the graviton mass term is droppsed by setting $(m =0)$, it reduces to the Schwarzschild-AdS black hole for $k = +1$~\cite{Cai2015,PVMassIV}. We also note that the choice of the reference metric with graviton mass terms leave the solution with a Lorentz-breaking property~\cite{Vegh}. It should also be mentioned that 
 the vacuum solution (obtained by setting $m_0 = 0$ in~\eqref{Y(r)}) is not an AdS space unless graviton mass is zero, i.e., $m=0$. 
\vskip 0.5cm

\noindent
The thermodynamic quantities are given in terms of the horizon radius $r_+$, which is the largest positive root of $Y(r_+)=0$. Especially, the temperature $T$, mass $M$ and entropy $S$ of the black hole are given as~\cite{Cai2015,PVMassIV}: 
\begin{eqnarray}
T &=& \frac{Y'(r_+)}{4\pi}= \frac{k}{4\pi r_{+}}-\frac{r_{+}\Lambda }{4\pi
}  +\frac{m^{2}}{4\pi r_{+}}\left(
c_0c_{1}r_{+}+c_{2}c_0^{2}\right) \ ,
\label{eq:Temp} \\
 M &=&\frac{m_{0}}{2 } = \frac{r_+}{2} \left(k -\frac{\Lambda}{3}r_+^2  + m^2(\frac{c_0 c_1}{2}r_+ + c_0^2c_2)\right)\ ,  \label{eq:Mass} \\
S &=&\pi r_{+}^{2} \ .
\end{eqnarray}
\noindent
In the extended thermodynamics approach, the pressure is provided by a dynamical cosmological constant\footnote{A scenario where cosmological constant is varying can occur in a gravity theory under various circumstances. For instance, in a situation where the theory is embedded in a larger set up consisting of other matter sectors, with possible origins in string theory. A dynamical cosmological constant can arise as a vev of a scalar field in a high energy theory under dimensional reduction, where there are several dynamical scalar fields with their respective potentials~\cite{Aharony:1999ti}. We do not pursue these aspects here.}, as $p=-\frac{\Lambda}{8\pi}=\frac{3}{8\pi l^2}$ in four dimensions, with $l$ being the AdS radius. The thermodynamic conjugate of $p$ is the thermodynamic volume $V$. In this set up one identifies the mass $M$ of the black hole with the enthalpy $H$~\cite{Kastor:2009wy}, leading to the first law of 
black hole thermodynamics given as~\cite{Cai2015,PVMassIV}:
\begin{equation}
dM=TdS+Vdp+\mathcal{C}_{1}dc_{1} \ ,  \label{1stlaw}
\end{equation}
where 
\begin{eqnarray}
V &=&\left( \frac{\partial M}{\partial p}\right) _{S,c_{1}}=\frac{4\pi}{3} r_+^3 \ ,  \label{Vol} \\
\mathcal{C}_{1} &=&\left( \frac{\partial M}{\partial c_{1}}\right)
_{S,p}=\frac{c_0m^{2}r_{+}^{2}}{4 } \ .
\label{C1} 
\end{eqnarray}
The specific heat at constant volume $C_V$, and at constant pressure $C_p$ are found to be~\cite{Hendi:2017bys}:  
 \begin{equation} \label{Cv}
 	C_V=T\Big(\frac{\partial S}{\partial T}\Big)_V = 0 \ ; \, \, \,   C_p= T\Big(\frac{\partial S}{\partial T}\Big)_p=  2S\Bigg(\frac{8pS^2+S(k+m^2c_0^2 c_2)  + \frac{m^2c_0 c_1 S^{3/2}}{\sqrt{\pi}}}{8pS^2 -S(k+m^2c_0^2 c_2)  } \Bigg) \ .
 \end{equation}

\subsection{Phase structure} \label{phase}

\begin{figure}[h!]
	{\centering
		\subfloat[]{\includegraphics[width=3in]{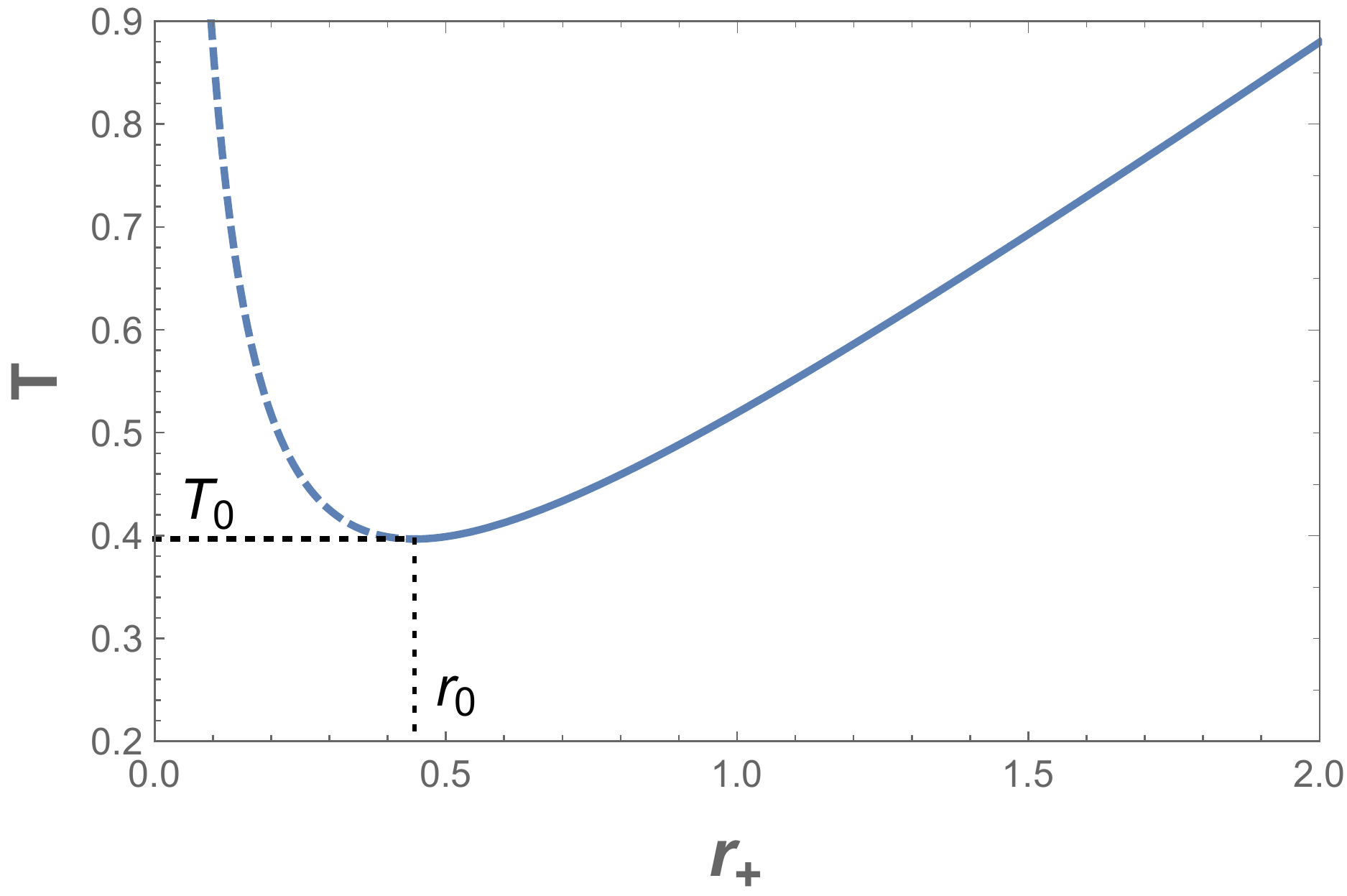}\label{Fig:T_r_plot}}\hspace{0.5cm}	
		\subfloat[]{\includegraphics[width=3in]{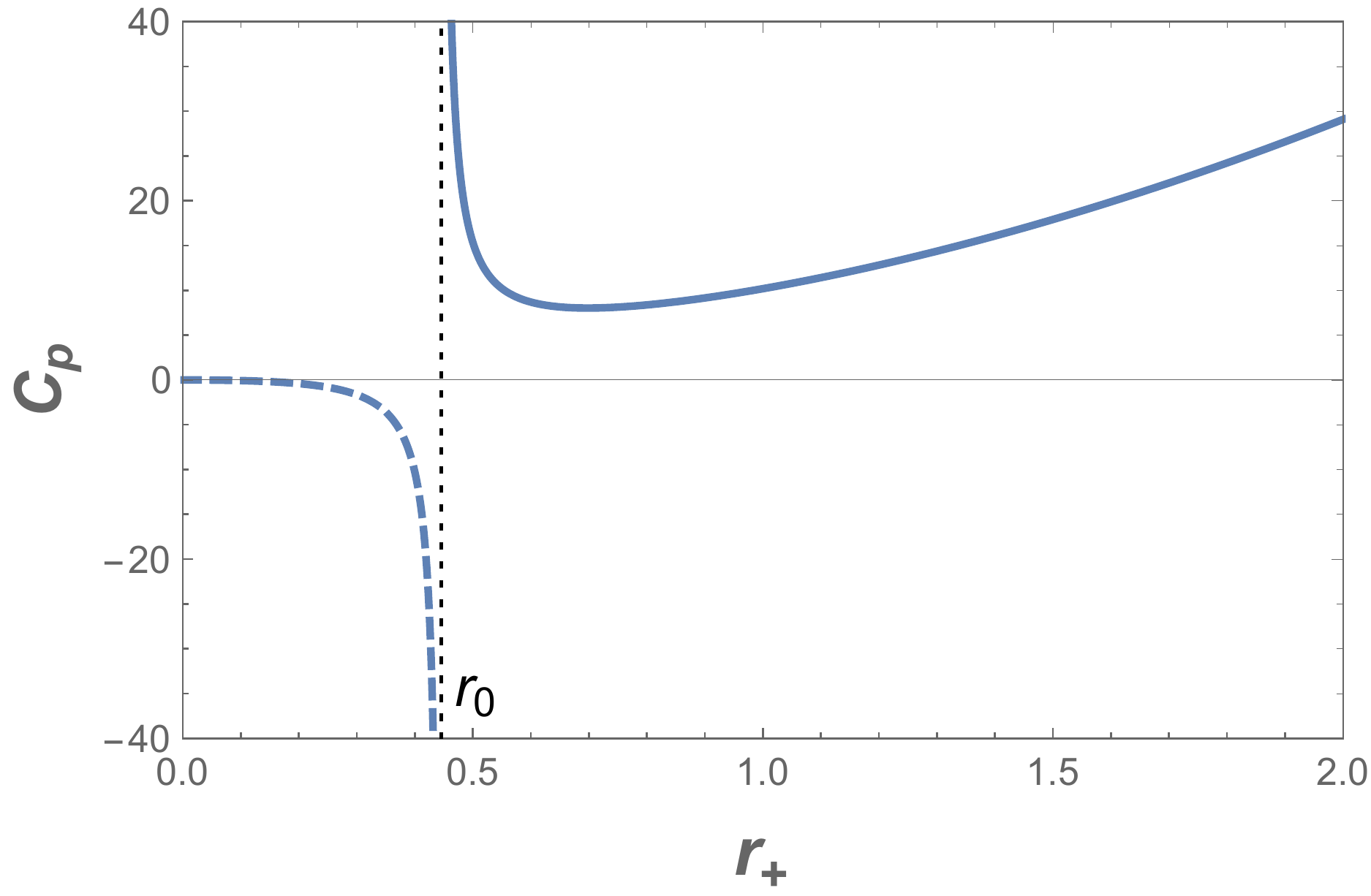}\label{Fig:Cp_r_plot}}

		\caption{\footnotesize For  the topological AdS black holes in massive gravity (a) Temperature $T$ versus horizon radius $r_+$ plot, indicating the existence of minimum temperature $T_0$ at $r_0$, when  $\epsilon \equiv (k+m^2 c_2 c_0^2) > 0 $. For $T < T_0$, no black holes can exist, except the vacuum phase. For $T > T_0$, there exist small (shown with dashed blue curve) and large (shown with solid blue curve) black holes. (b) Specific heat $C_p$ versus $r_+$ plot, indicating  the negative $C_p$ for small black holes (dashed blue curve), positive $C_p$ for large black holes (solid blue curve), and $C_p$ diverges for the black hole with horizon radius $r_0$.   (Here, the parameters $k=-1, \ m=c_0 =1, \ c_1=0.5, \ c_2=2, \ p=0.2$, are used. Similar behavior is observed for other topologies.).} 
	}
\end{figure}

\begin{figure}[h!]
	{\centering
		\subfloat[]{\includegraphics[width=3in]{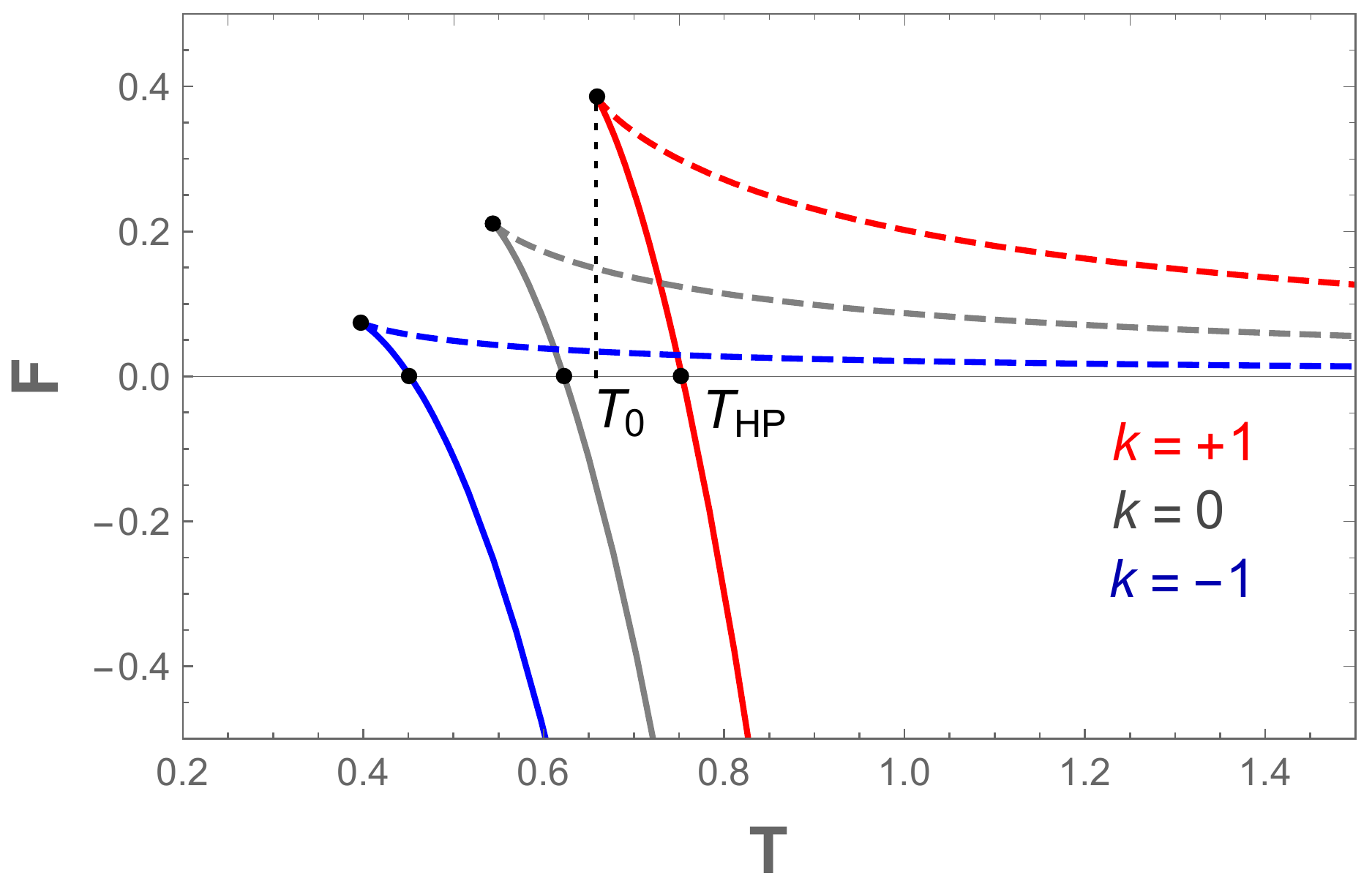}\label{Fig:F_T_plot}}\hspace{0.5cm}	
		\subfloat[]{\includegraphics[width=3in]{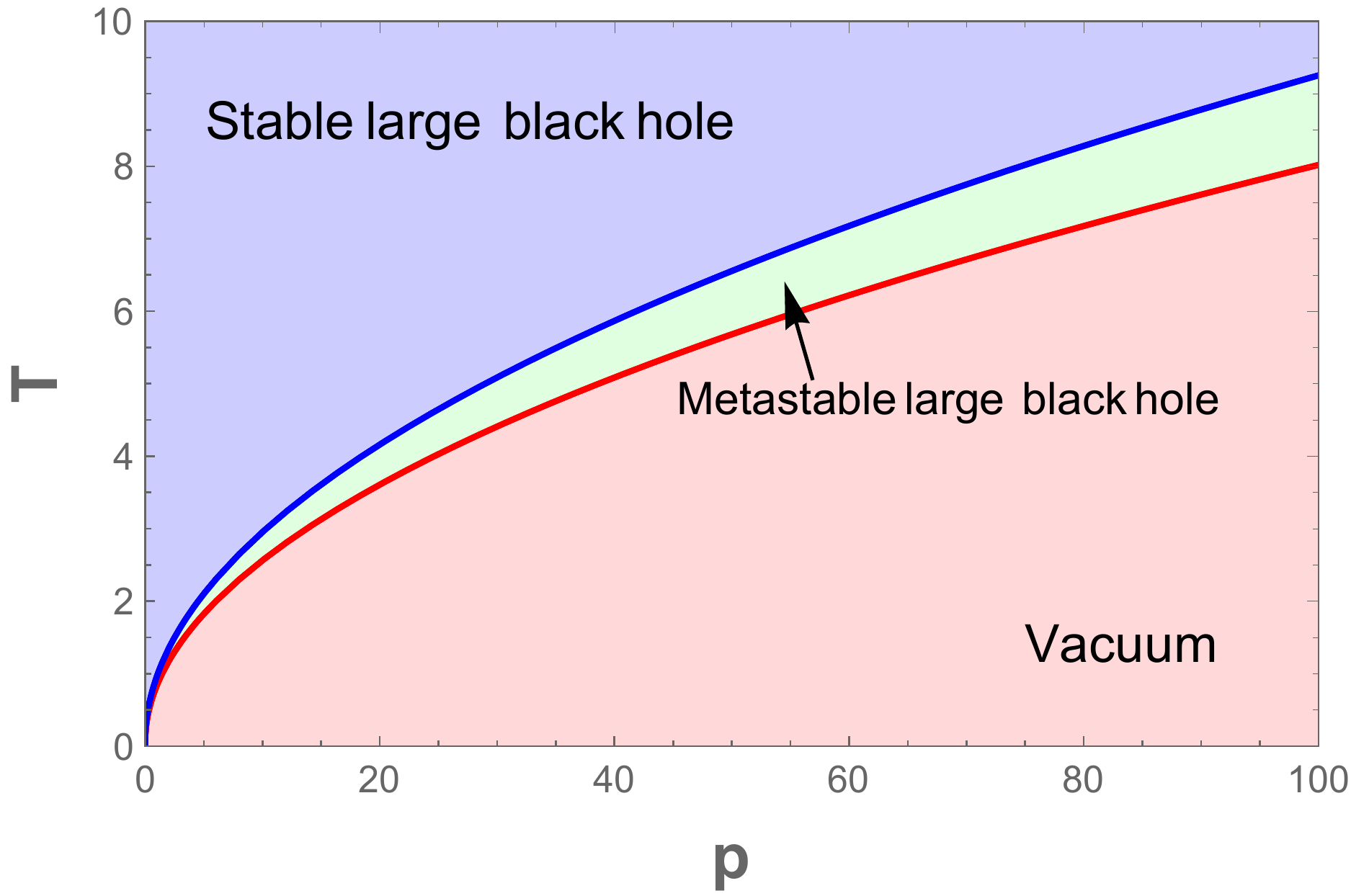}\label{Fig:T_p_plot}}

		\caption{\footnotesize For  the topological AdS black holes in massive gravity (a) Free energy $F$ versus Temperature $T$ plot at various horizon topologies $k$ for a fixed pressure $p= 0.2$. Dashed curve  indicates the small black hole branch, solid curve indicates the large black hole branch, $T_0$ is the minimum temperature and $T_{HP}$ is the Hawking-Page phase transition temperature.  (b) Phase structure in $T-p$ plane for horizon topology $k=-1$ (Similar phase structure is observed for other topologies). The red  and blue  curves respectively correspond to the black hole minimum temperature $T_0$ and the HP phase transition temperature $T_{HP}$.   (Here, the parameters $  m=c_0 =1, \ c_1=0.5, \ c_2=2$, are used.).} 
	}
\end{figure}

From the expression of Hawking temperature~\eqref{eq:Temp}, when 
\begin{equation}\label{epsilon}
\epsilon \equiv (k +m^2c_2 c_0^2),
\end{equation} 
is greater than zero, one can see that there exists a minimum temperature $T_0$ of the black hole, given by~\cite{Cai2015}
\begin{equation}
T_{min} = T_0 = \sqrt{\frac{2 \epsilon p }{\pi}}  + \frac{m^2 c_0 c_1}{4\pi},
\end{equation}
with horizon  radius $r_0 = \sqrt{\frac{\epsilon}{8\pi p}}$.  For  $T > T_0$, there exist a pair of black holes, small ($r_+ < r_0$) and large ($r_+ > r_0 $), as shown in Fig.~\ref{Fig:T_r_plot}. The small black holes are thermodynamically unstable due to negative specific heat $C_p$, whereas  the large black holes are stable with positive $C_p$, while $C_p$ diverges for the black hole with horizon radius $r_0$ (See Fig.~\ref{Fig:Cp_r_plot}). 
\\
 \noindent As shown in Fig.~\ref{Fig:F_T_plot}, for $T < T_0$, no black holes can exist, except the vacuum phase characterized by vanishing free energy $F$. The large black holes  are metastable having positive free energy when $T_0 < T < T_{HP}$, while at $T=T_{HP}$, the free energy of the large black hole equals to the vacuum free energy, where the Hawking-Page (HP) transition happens between the vacuum phase and large black hole phase. 
 For $T> T_{HP}$, the large black hole phase having negative free energy is globally stable than vacuum phase~\cite{Cai2015}.
\vskip 0.4cm
 \noindent The expressions for the free energy $F$ and the Hawking-Page phase transition temperature $T_{HP}$ are given by~\cite{Cai2015}; 
\begin{eqnarray}
F &=& M -TS = \frac{r_+}{4}\Big( \epsilon - \frac{8\pi p}{3}r_+^2 \Big), \\
 T_{HP} &=& \sqrt{\frac{8\epsilon p}{3\pi}} +\frac{m^2c_0 c_1}{4\pi}. \label{eq:Thp_massive}
\end{eqnarray}
The Hawking-Page phase transition happens at the horizon radius $r_{HP} = \sqrt{\frac{3\epsilon}{8\pi p}}$. The phase structure, as shown in Fig.~\ref{Fig:T_p_plot}, is identical to that of the  Schwarzschild-AdS black holes with spherical topology $(k = +1)$ of massless graviton case~\cite{Belhaj:2015hha,Wei:2020kra}. 
Thus, provided $\epsilon > 0$, the presence of massive graviton  admits the Hawking-Page transition for the black holes with topology flat  $(k=0)$ and  hyperbolic $(k=-1)$ as well, unlike the case of massless graviton, where only the black holes with spherical topology $(k=+1)$ can under go Hawking-Page transition~\cite{Birmingham:1998nr,Cai2015}. \\
\begin{figure}[h!]
	{\centering
		\subfloat[]{\includegraphics[width=3in]{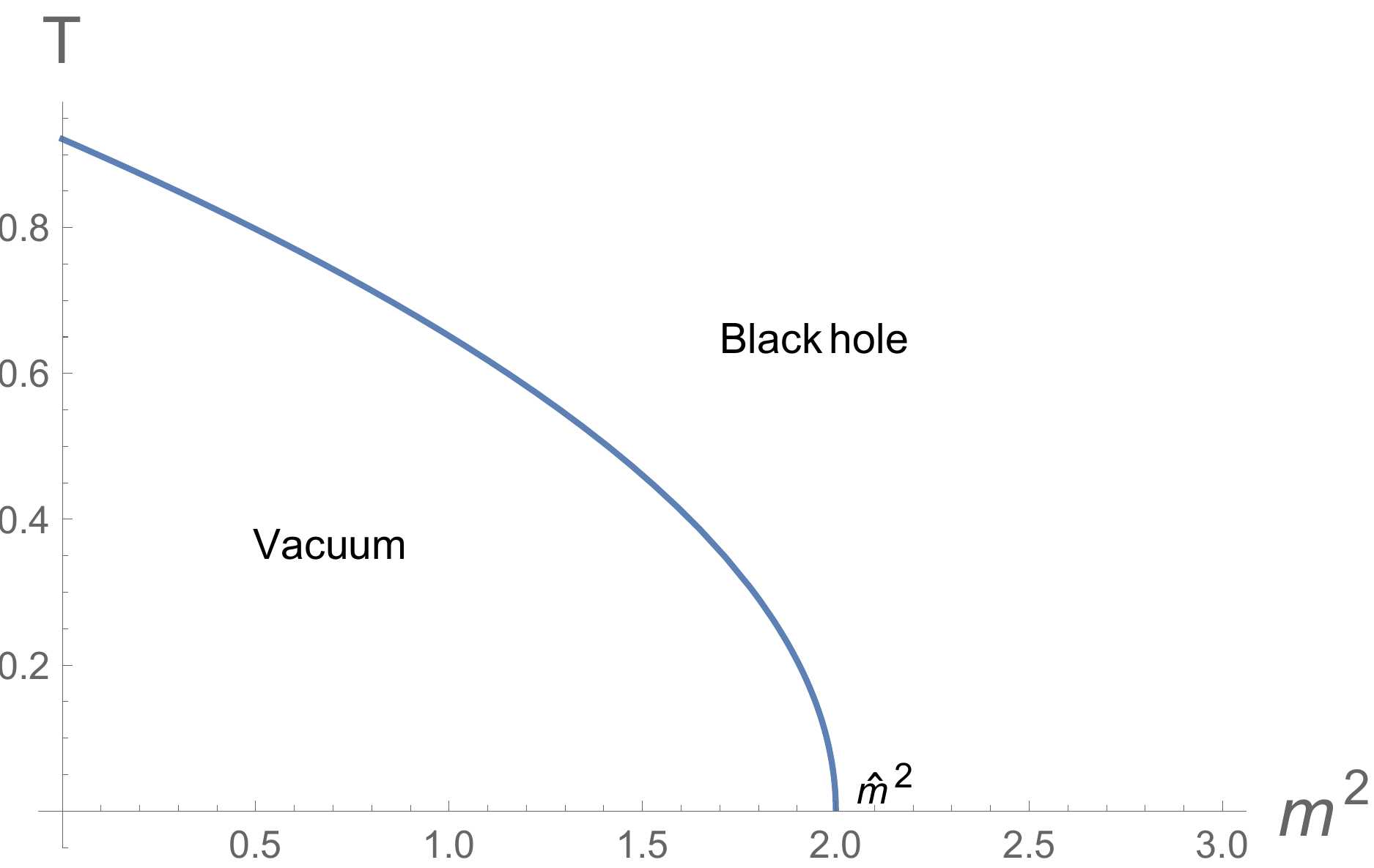}\label{zeroHP1}}\hspace{0.5cm}	
		\subfloat[]{\includegraphics[width=3in]{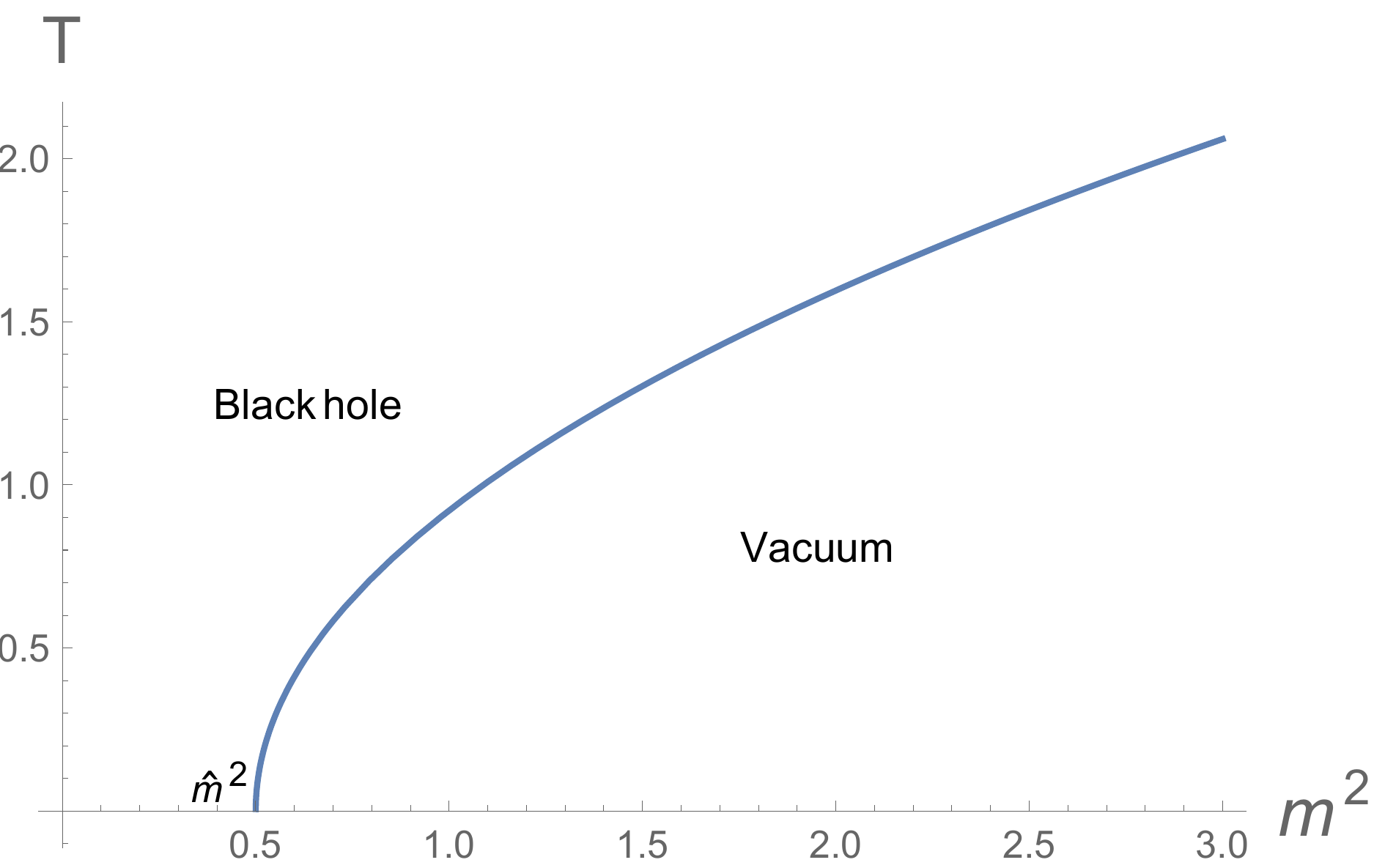}\label{zeroHP2}}

		\caption{\footnotesize The phase diagram in the case of massive coefficient $c_1 = 0$, showing that the Hawking-Page transition temperature $T_{HP} (m^2)$  vanish at $ m^2= \hat{m}^2 = -k/(c_2 c_0^2)$  for the black holes with a) spherical horizon topology (Here, the parameters  $p=c_0=1, c_2=-1/2 $, are used matching~\cite{Adams:2014vza}), b) hyperbolic horizon topology  (Here, the parameters  $p=c_0=1, c_2=2 $, are used). The first order phase transition line terminates in a second order transition point at zero temperature and $ m^2 = \hat{m}^2$}
	}
\end{figure}
\vskip 0.1cm
\noindent
From eqn. (\ref{eq:Thp_massive}), one notes the possibility of having the HP transition at zero temperature for specific choice of parameters. Such a phenomenon in case of black holes with spherical topology, where the dual field theory also undergoes deconfinement transition at zero temperature  was already noted earlier in a specific holographic massive gravity model~\cite{Adams:2014vza} (see the caption of figure-(\ref{zeroHP1}) for the choice of parameters).  The graviton mass was argued to encode the rate of momentum dissipation in the dual field theory and drive the critical temperature for deconfinement to zero, at a new length scale set by $l_m \approx 1/m$. This length scale determines the distance which the particles travel before shedding momentum.
For generic horizon topology in the bulk, the point where the HP transition temperature goes to zero for different possible choices of parameters and graviton mass can be studied. For instance, one can consider $ c_1 < 0 $ where naively the HP transition temperature does approach zero, but the vacuum solution for this case contains horizons and hence is not suitable (arguments are similar to the ones given in~\cite{Adams:2014vza}). The case with $c_1=0$ is more interesting as seen from figures-(\ref{zeroHP1}) and (\ref{zeroHP2}) for black holes with spherical and hyperbolic topology respectively. The curves indicate the line of first order transition separating the vacuum and black hole phases, with HP transition driven to zero temperature at a critical value of graviton mass, $\hat{m}^2 = -k/(c_2 c_0^2)$. In the field theory, as the critical value $\hat{m}^2$ is crossed ($m^2 > \hat{m}^2$ for spherical topology and $m^2 < \hat{m}^2$ for hyperbolic topology), the momentum dissipation effects grow strong with the length scale $l_m$ falling below the confinement scale, and the system effectively deconfines. All these are interesting effects and it is important to understand the behavior of microstructures in some of these special limits. The plan of the next two sections is to gather empirical information from thermodynamic curvature, especially as $m^2$ approaches $\hat{m}^2$,  as well as explore the geometry around the HP transition point in general, in novel near horizon limits.

\section{Thermodynamic curvature at the HP transition point} \label{3}

\noindent
Thermodynamic or Ruppeiner geometry can be thought of as a macroscopic approach to gain understanding about the microscopic aspects of a thermodynamic system, where a thermodynamic curvature scalar carries much information about the type of interactions. Thermodynamic geometry has its roots in fluctuation theory, which starts from studying the inverse of Boltzman entropy formula, namely:
\begin{equation}\label{omega1}
\Omega = e^{\frac{S}{k_B}}\, ,
\end{equation}
where $\Omega$ stands for the number of microstates, $S$ is the entropy and $k_B$ is the Boltzmann constant.
One starts by considering a large thermodynamic system $I_0$ in equilibrium, that consists in it a sub-system $I$, with the later having, say, two independent fluctuating coordinates, $x^i$ where $i=1,2$. 
The probability $P(x^1,x^2)$ of having the state of the system between $(x^1,x^2)$ and $(x^1 + dx^1,x^2+dx^2)$ can straightforwardly be related to the number of microstates $\Omega$ from eqn. (\ref{omega1}).  In this situation, the second law of thermodynamics states that the pair $(x^1,x^2)$  are frozen on the values which maximize the entropy $S=S_{\text{max}}$. In other words, the pair $(x^1,x^2)$ actually describes thermodynamic fluctuations around this maximum. One can now expand the entropy about this maximum up to second order, which shows that the probability is~\cite{Ruppeiner:1995zz}:
\begin{equation}
P(x^1,x^2) \propto e^{-\frac{1}{2}\, \Delta l^2} \, .
\end{equation}
Now,
\begin{equation}\label{distance}
\Delta l^2 \, = -\frac{1}{k_B}\, \frac{\partial^2 S}{\partial x^i\partial x^j}\, \Delta x^i \Delta x^j,
\end{equation}
stands for the line element, which is a measure of the thermodynamic distance between two neighborhood fluctuating states. If the distance between these states is shorter, the more probable is the fluctuation between them. Based on various studies of the thermodynamic curvature $R$, following from the metric in eqn. (\ref{distance}) for various fluid/gas systems, such as ideal and van der Waals systems, Fermi/Bose systems including quantum gases, the empirical understanding gained can be summarized as follows: A negative (positive) sign of $R$ indicates that attractive (repulsive) type interactions are dominant in the system. Furthermore, a larger negative (positive) value of curvature signifies that the system is less (more) stable, which further points towards the stability of Bose (Fermi) type systems
~\cite{RuppeinerRMP,Ruppeiner2,Ruppeiner3,Janyszek:1989zz,Janyszek1990}. The divergences of $R$ indicate the presence of critical points and vanishing of curvature is suggestive of balance of repulsive and attractive interactions, namely, a non-interactive situation.\\

\noindent
To obtain the thermodynamic curvature, the line element in eqn. (\ref{distance}) needs to be computed for a choice of fluctuation coordinates. In the current situation where entropy is clearly the key thermodynamic quantity, the line element can be obtained by starting from the internal energy $U=U(S,V)$. We have the first law $dU = T dS - P dV$, which can be written as:
\begin{equation}
dS = \frac{1}{T} dU + \frac{P}{T} dV \, ,
\end{equation}
with $V$ representing the thermodynamic volume, conjugate to pressure $P = - \Lambda/{8\pi G}$~\cite{Kastor:2009wy,Dolan:2011xt,Cvetic:2010jb}. We now take the fluctuating variables to be  $(x^1=T,x^2=V)$, in which case the line element can be written explicitly as~\cite{Landau,Wei:2019uqg,Wei:2019yvs}:
\begin{equation} 
dl^2_{U} = \frac{1}{T} \left(\frac{\partial S}{\partial T} \right) \Bigr|_V dT^2 - \frac{1}{T} \left(\frac{\partial P}{\partial V} \right) \Bigr|_T dV^2\, ,
\end{equation}
which can be also be written in a useful form as:
\begin{equation}\label{metricU}
dl^2 = \frac{C_V}{T^2}dT^2 - \frac{1}{T}\Big(\frac{\partial p}{\partial V}\Big)_T dV^2.
\end{equation} 
At this stage, the metric in eqn. (\ref{metricU}) is quite general and any thermodynamic system (non necessarily a black hole) can be studied by computing the associated curvature from it. Microstructures of van der Waals liquid-gas system studied from the curvature coming from the line element in eqn. (\ref{metricU}) reveal a completely coherent and clear picture of the nature of interactions~\cite{Wei:2019uqg,Wei:2019yvs}. For some specific case of static black holes in AdS~\cite{Chamblin:1999tk,Kubiznak:2012wp}, additional normalization of the thermodynamic curvature may be required to extract the nature of interactions of microstructures~\cite{Wei:2019uqg,Wei:2019yvs}. This can be seen from the fact $\left(\frac{\partial S}{\partial T} \right)_V= C_V/T$ is zero from equation (\ref{Cv}), translating to the metric being non-invertible. One can deal with this situation by momentarily assuming $C_V$ to a small non zero quantity and as the $k_B \rightarrow 0^{+}$ limit~\cite{Wei:2019uqg,Wei:2019yvs}. The curvature $R$ can now be computed and $C_V$ comes to be a overall multiplicative constant in it. One can rescale to remove the dependence on $C_V$ and define a new curvature as $R_N = R \, C_V$, which carries the same empirical information as $R$ and gives consistent results for black holes~\cite{Wei:2019uqg,Wei:2019yvs,DasBairagya:2020rqy}. \\

\begin{figure}[h!] 
 	{\centering
 		\subfloat[]{\includegraphics[width=2in]{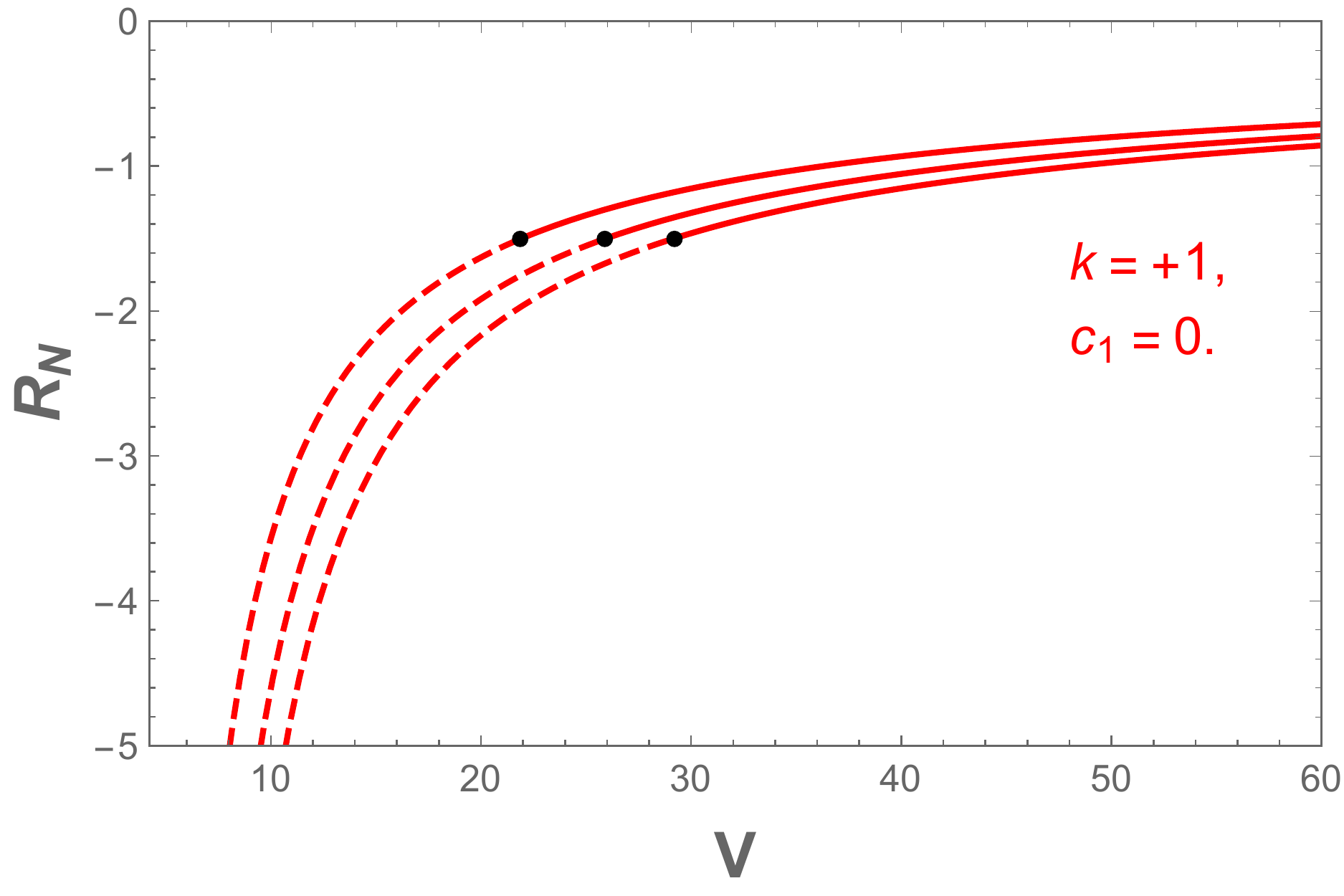}}\hspace{0.15cm}
 		\subfloat[]{\includegraphics[width=2in]{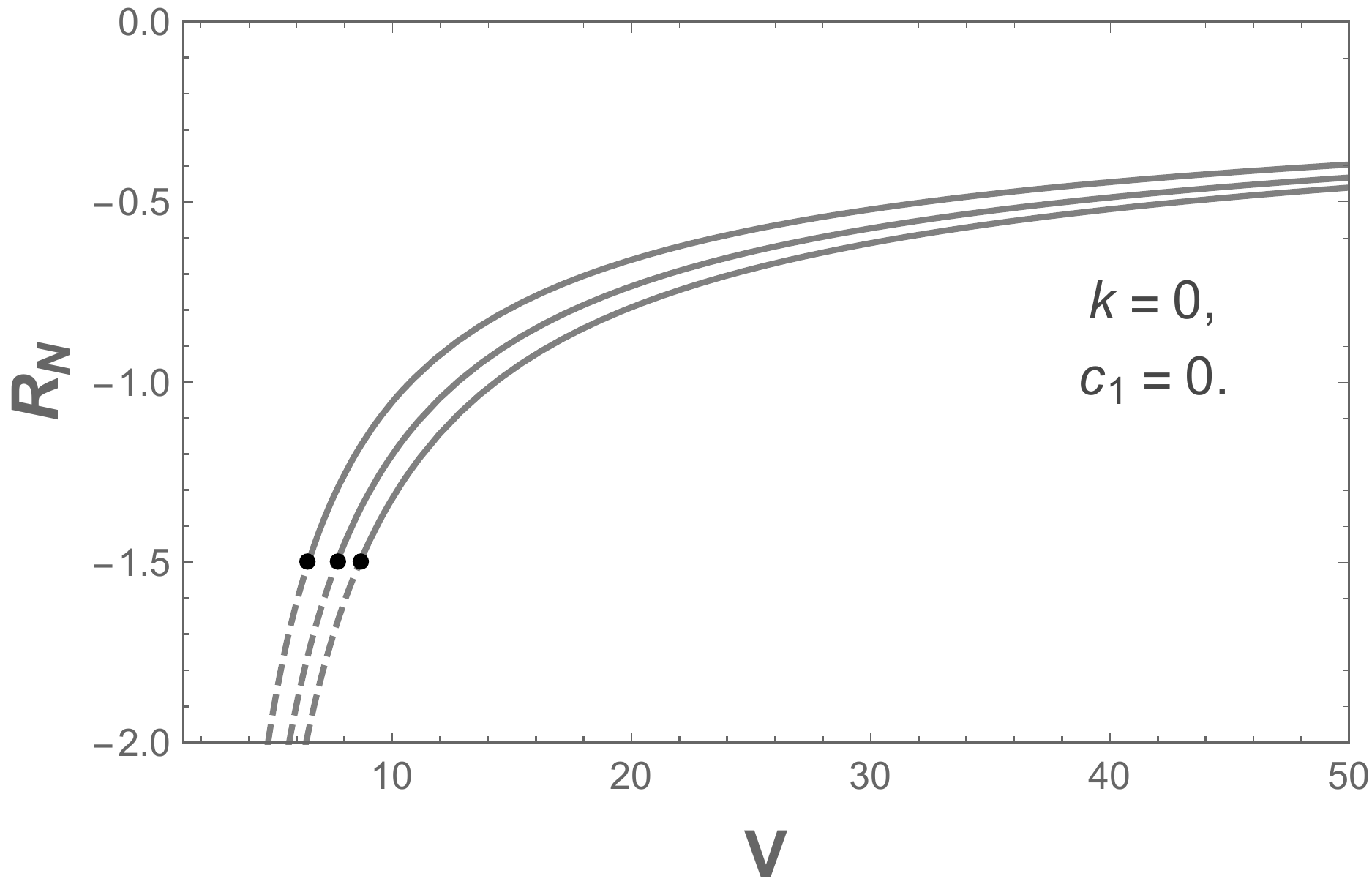}}\hspace{0.15cm}
 		\subfloat[]{\includegraphics[width=2in]{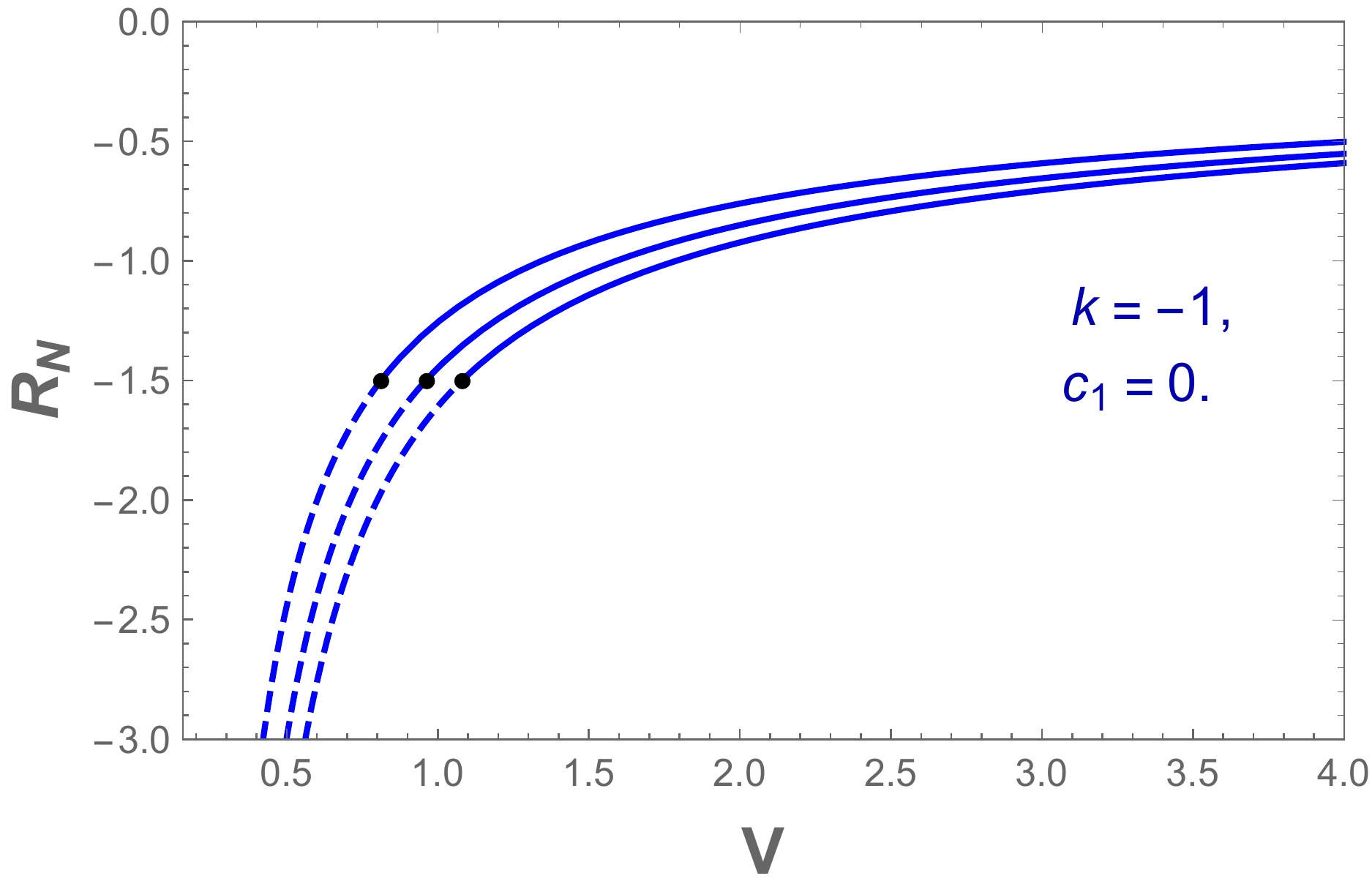}}\hspace{0.15cm}
 		\subfloat[]{\includegraphics[width=2in]{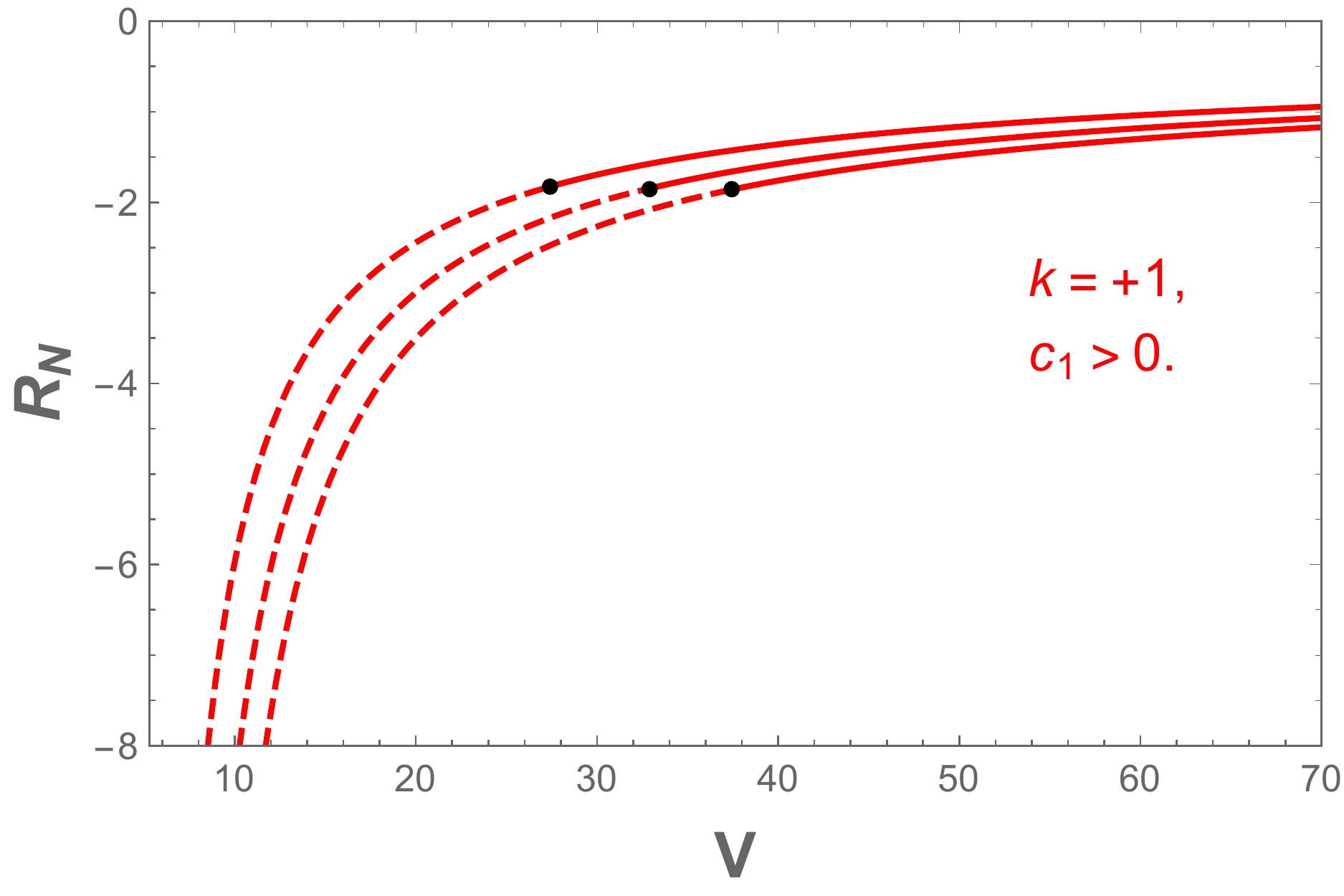}}\hspace{0.15cm}
 		\subfloat[]{\includegraphics[width=2in]{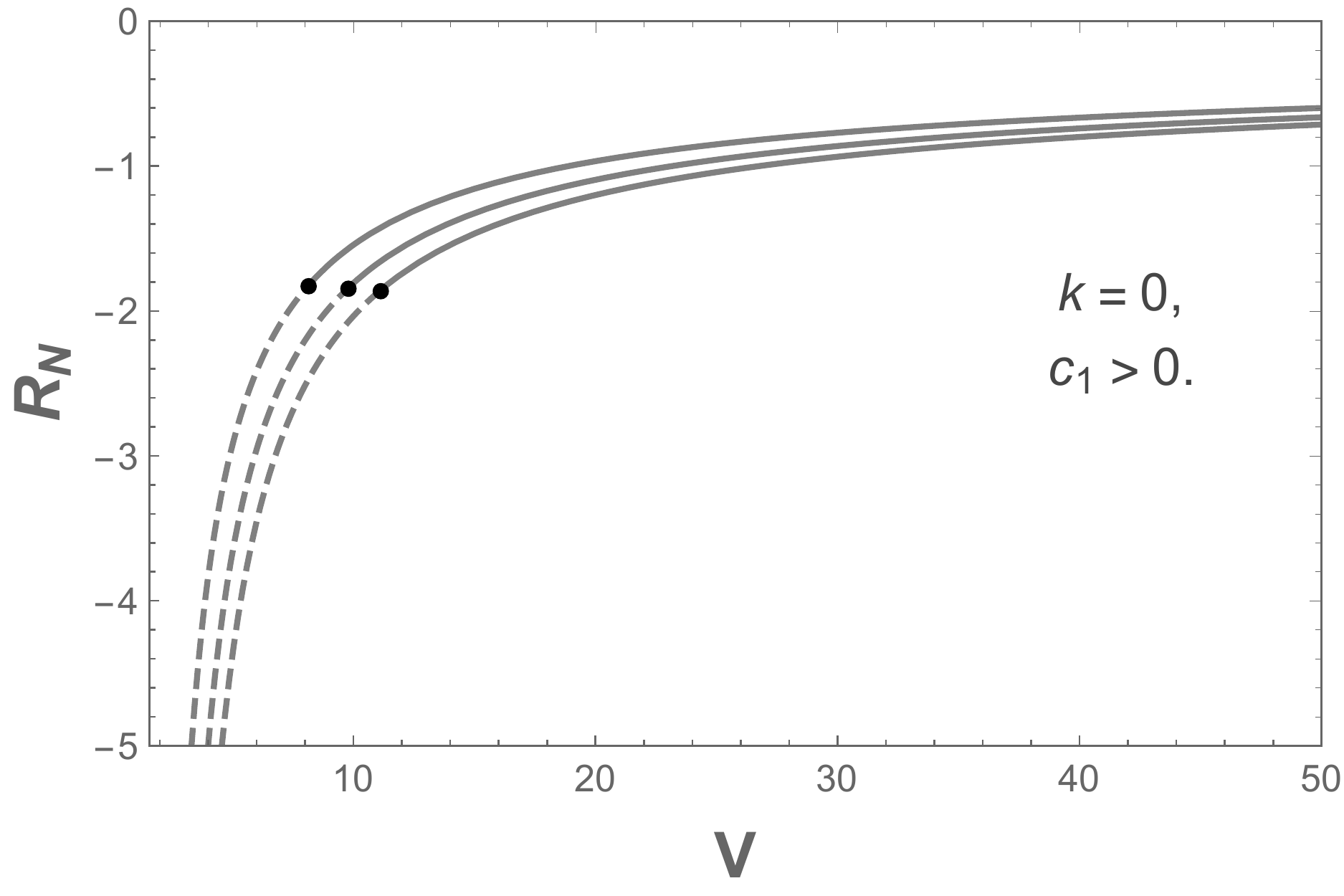}}\hspace{0.15cm}
 		\subfloat[]{\includegraphics[width=2in]{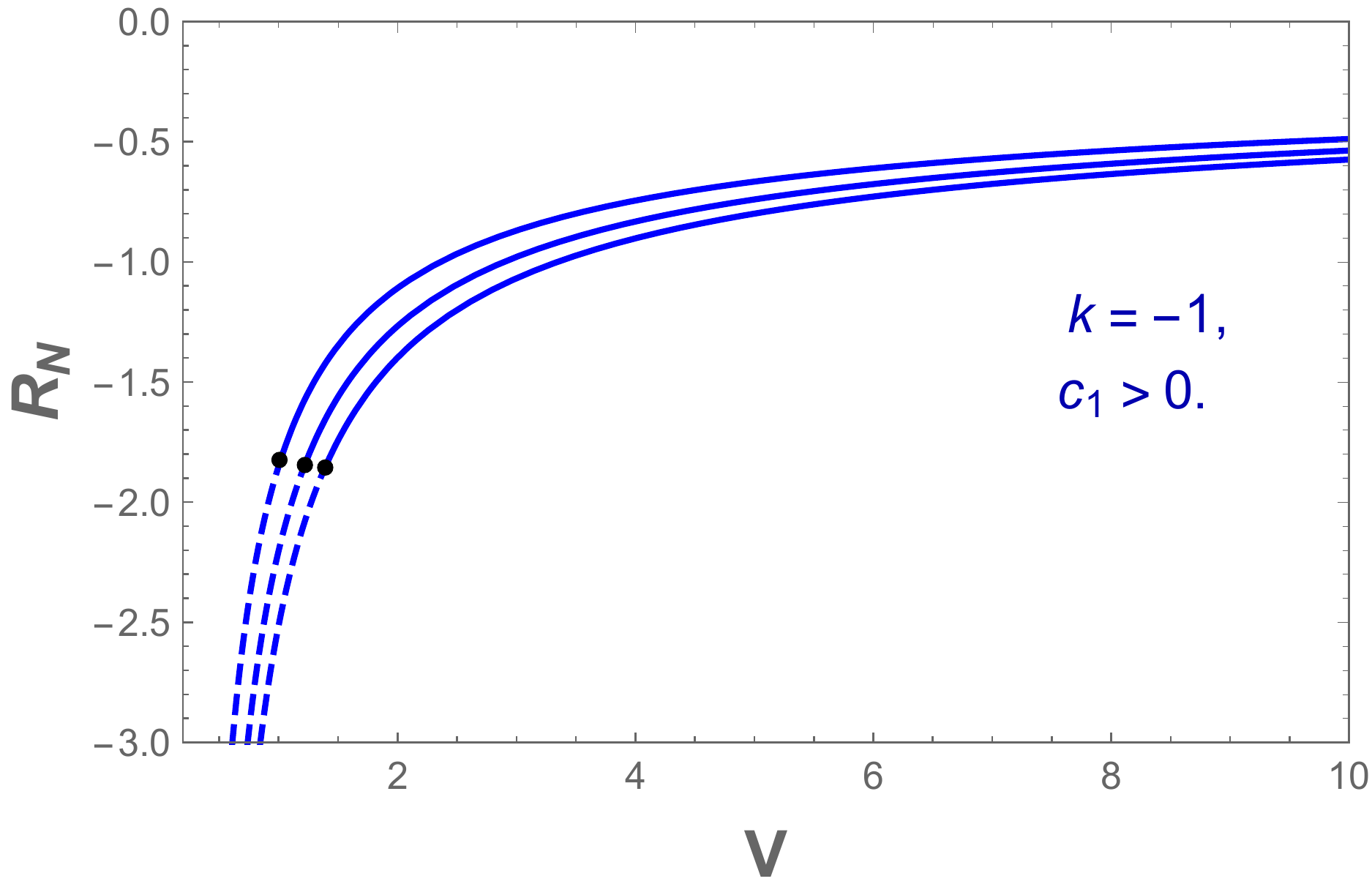}}				
	
 		\caption{\footnotesize In the case of massive coefficient $c_1 \geq 0$, the normalized scalar curvature $R_N$ as a function of thermodynamic volume $V$ for T = 0.5, 0.52, 0.55 from bottom to top, at various horizon topologies $k$. The dashed and  solid curves are for the metastable and
 			stable large black holes, respectively, while the black color dots represent HP transition points.  (Here, the parameters $m =c_0=1,\ c_1=0.5, \ c_2=2$, are used.). 
 	\label{fig:RN_c1>=0}	}  
 	}
 \end{figure}

\noindent
To proceed further, we use $p=-\frac{\Lambda}{8\pi}$ to invert the equation~\eqref{eq:Temp} and obtain the expression for equation of state $p(V, T)$ as: 
\begin{equation}
p  =  \frac{1}{8\pi}\Bigg\{\frac{(4\pi T-m^2c_0c_1)}{(\frac{3V}{4\pi})^{\frac{1}{3}}}-\frac{(k+m^2c_2c_0^2)}{(\frac{3V}{4\pi})^{\frac{2}{3}}}\Bigg\} \ . \label{eq of st} 
\end{equation}
Using the above expression in equation-(\ref{metricU}) and with a suitable normalization taking $C_V$ to be a constant as discussed above, on obtains:
\begin{equation}\label{eq:RN}
R_N = \frac{\Big( 2\epsilon (36\pi)^{\frac{1}{3}}  +3 c_0 c_1 m^2 V^{\frac{1}{3}}\Big)\Big( 2\epsilon (36\pi)^{\frac{1}{3}} -24\pi TV^{\frac{1}{3}} +3 c_0 c_1 m^2 V^{\frac{1}{3}}\Big)}{2\Big( 2\epsilon (36\pi)^{\frac{1}{3}} -12\pi TV^{\frac{1}{3}} +3 c_0 c_1 m^2 V^{\frac{1}{3}}\Big)^2}.
\end{equation}
Let us start by mentioning that in the limit where the graviton mass is set to zero, and for black holes with spherical topology $(k=+1)$, equation~\eqref{eq:RN} reproduces the result for thermodynamic curvature in~\cite{Wei:2020kra} and is also consistent with~\cite{Yerra:2020oph} in the appropriate limit. For the present case, $R_N$ diverges for the black hole with minimum temperature and this is consistent with the picture that the specific heat at constant pressure in equation-(\ref{Cv}) also diverges at this point. Since the small black hole branch is unstable,  we do not pursue this branch further.  For the large black hole branch, $R_N$ is finite and its sign  depends on the sign of the massive gravity coefficient $c_1$. 
 For  $c_1 \geq 0$, as shown in Fig.~\ref{fig:RN_c1>=0}, $R_N$ is negative  for the large black hole branch, indicating the dominance of attractive interactions for both metastable and stable large black holes, irrespective of horizon topology. \\

\begin{figure}[h!]
  	{\centering
  		\subfloat[]{\includegraphics[width=2in]{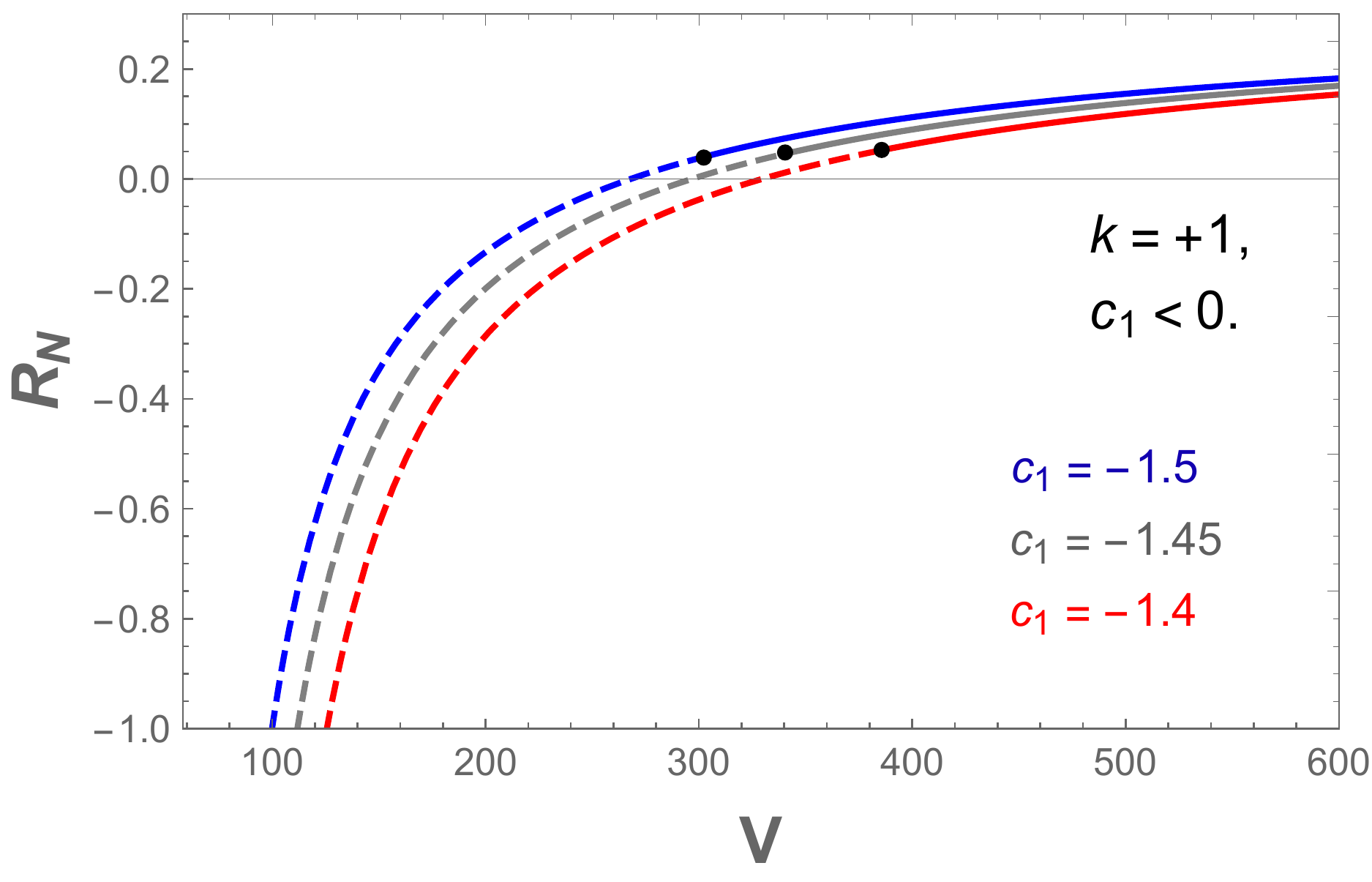}}\hspace{0.15cm}
  		\subfloat[]{\includegraphics[width=2in]{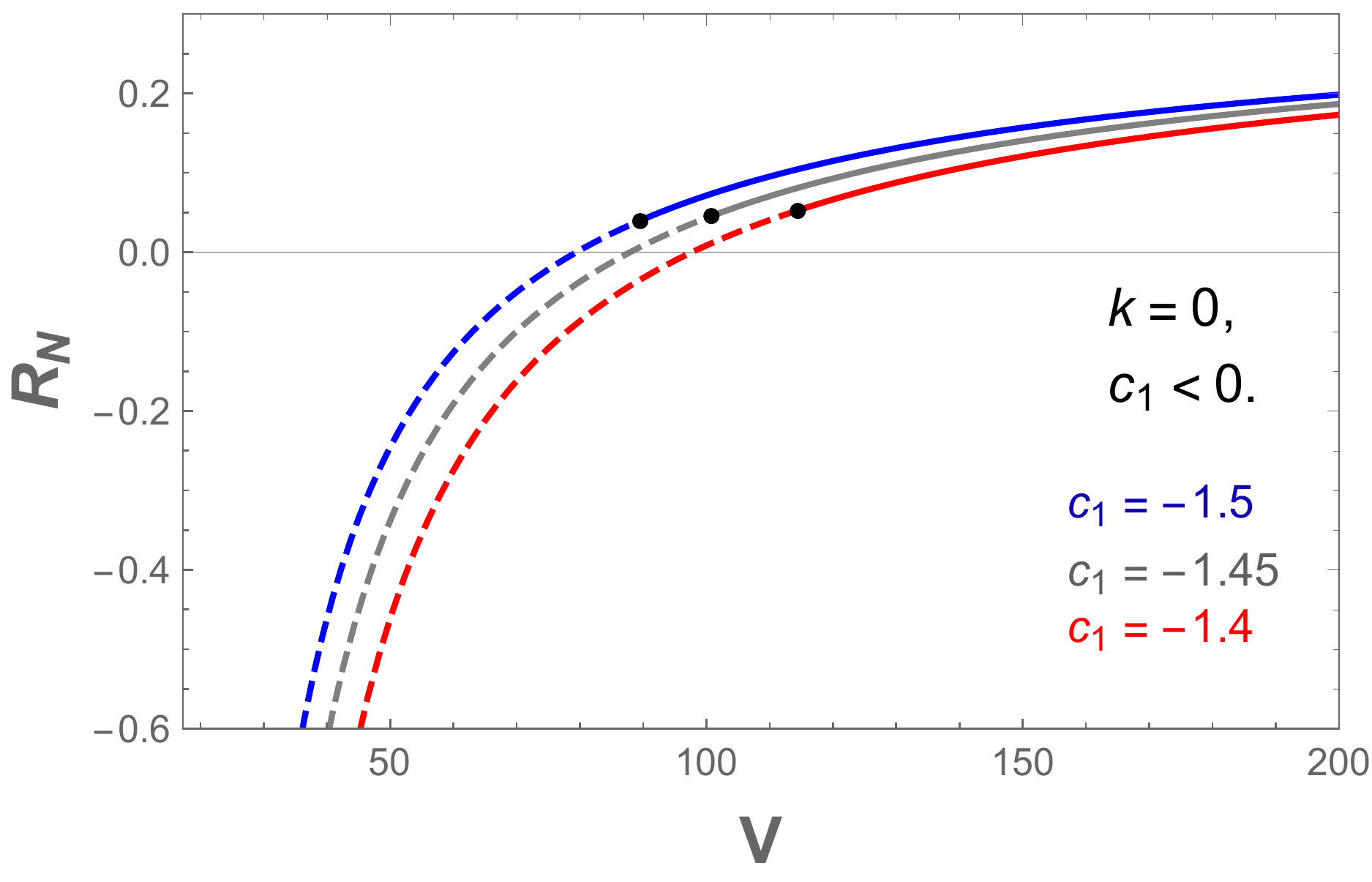}}\hspace{0.15cm}
  		\subfloat[]{\includegraphics[width=2in]{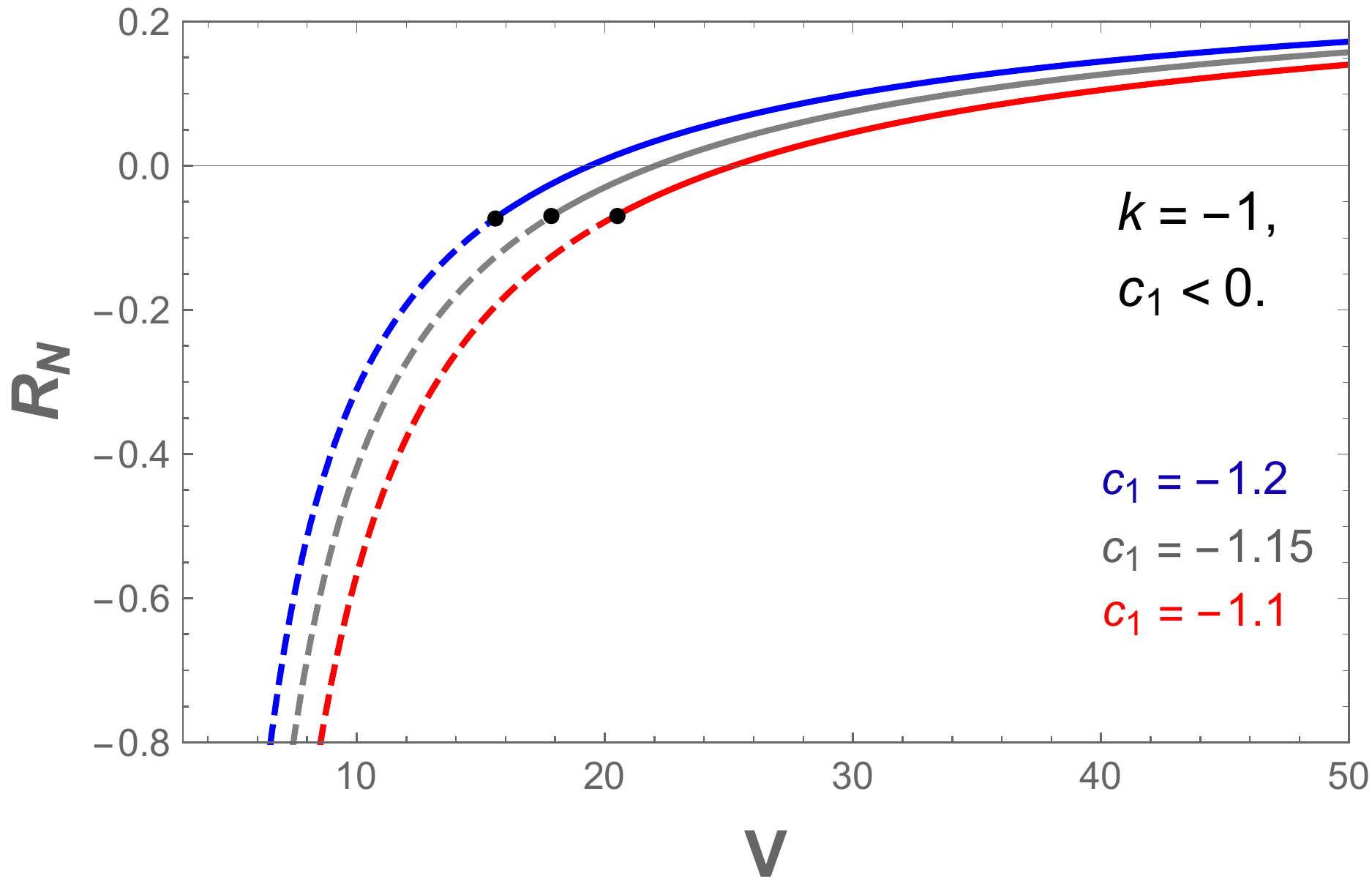}}
  	
  			\caption{\footnotesize In the case of massive coefficient $c_1 < 0$, the normalized scalar curvature $R_N$ as a function of thermodynamic volume $V$ for T = 0.1, 0.105, 0.11 from bottom to top, at various horizon topologies $k$. The dashed and  solid curves are for the metastable and
  				stable large black holes, respectively, while the black color dots represent HP transition points.  (Here, the parameters $m =c_0=1, \ c_2=2$, are used.).  \label{fig:RN_c1<0}	}  
  	}  
  \end{figure}

\begin{figure}[h!]
 	{\centering
 		\subfloat[]{\includegraphics[width=2in]{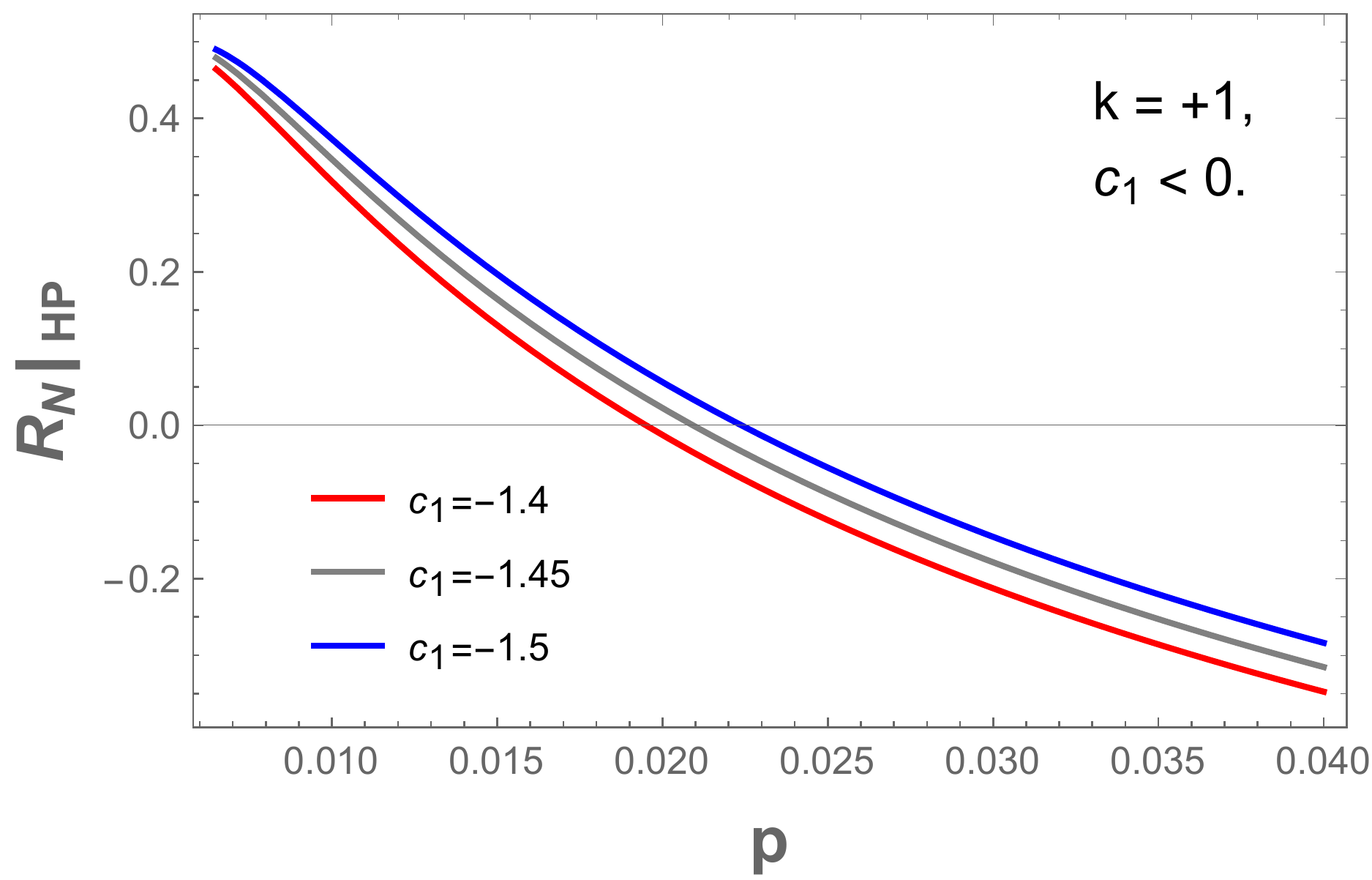}}\hspace{0.15cm}
 		\subfloat[]{\includegraphics[width=2in]{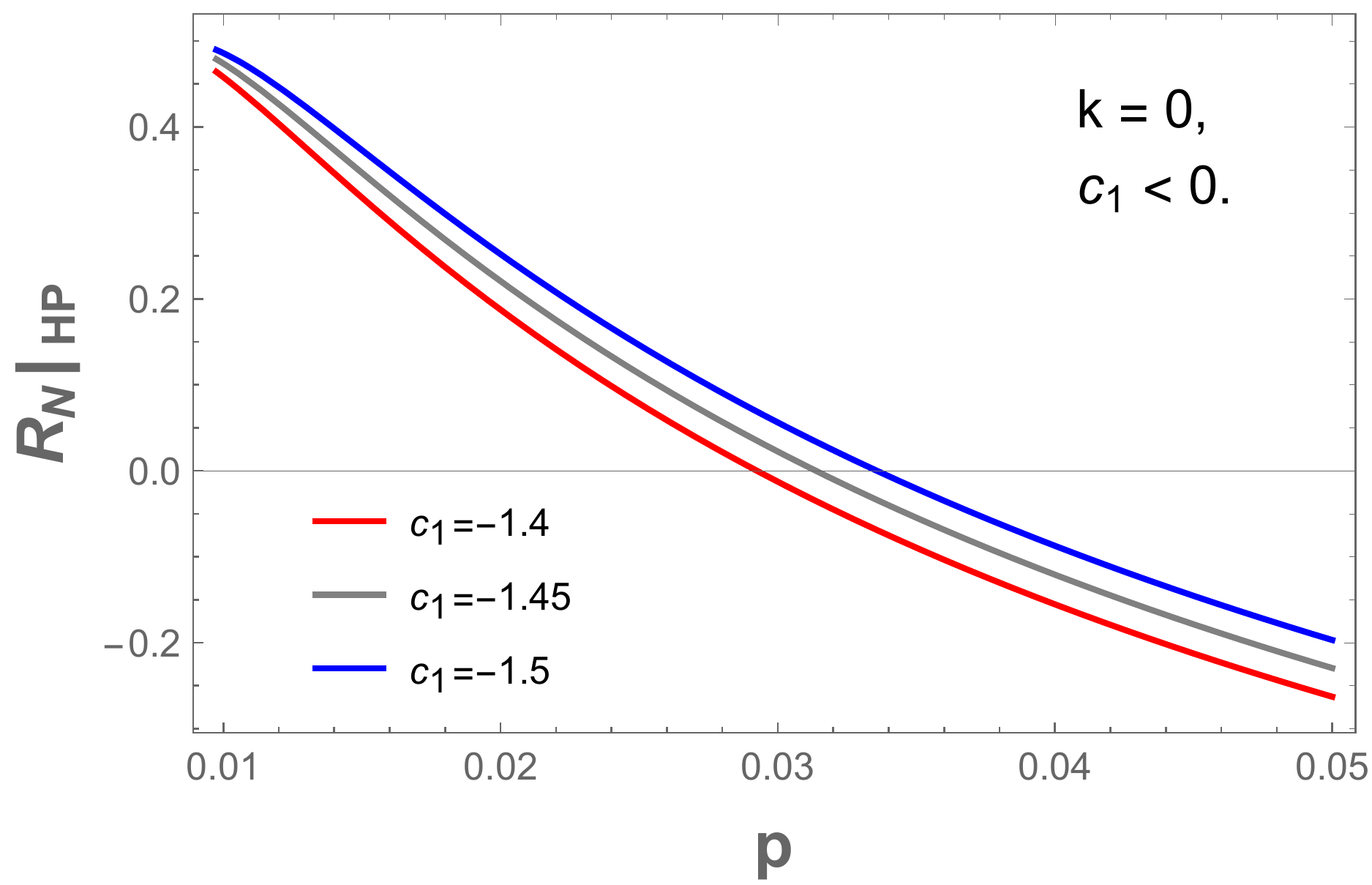}}\hspace{0.15cm}
 		\subfloat[]{\includegraphics[width=2in]{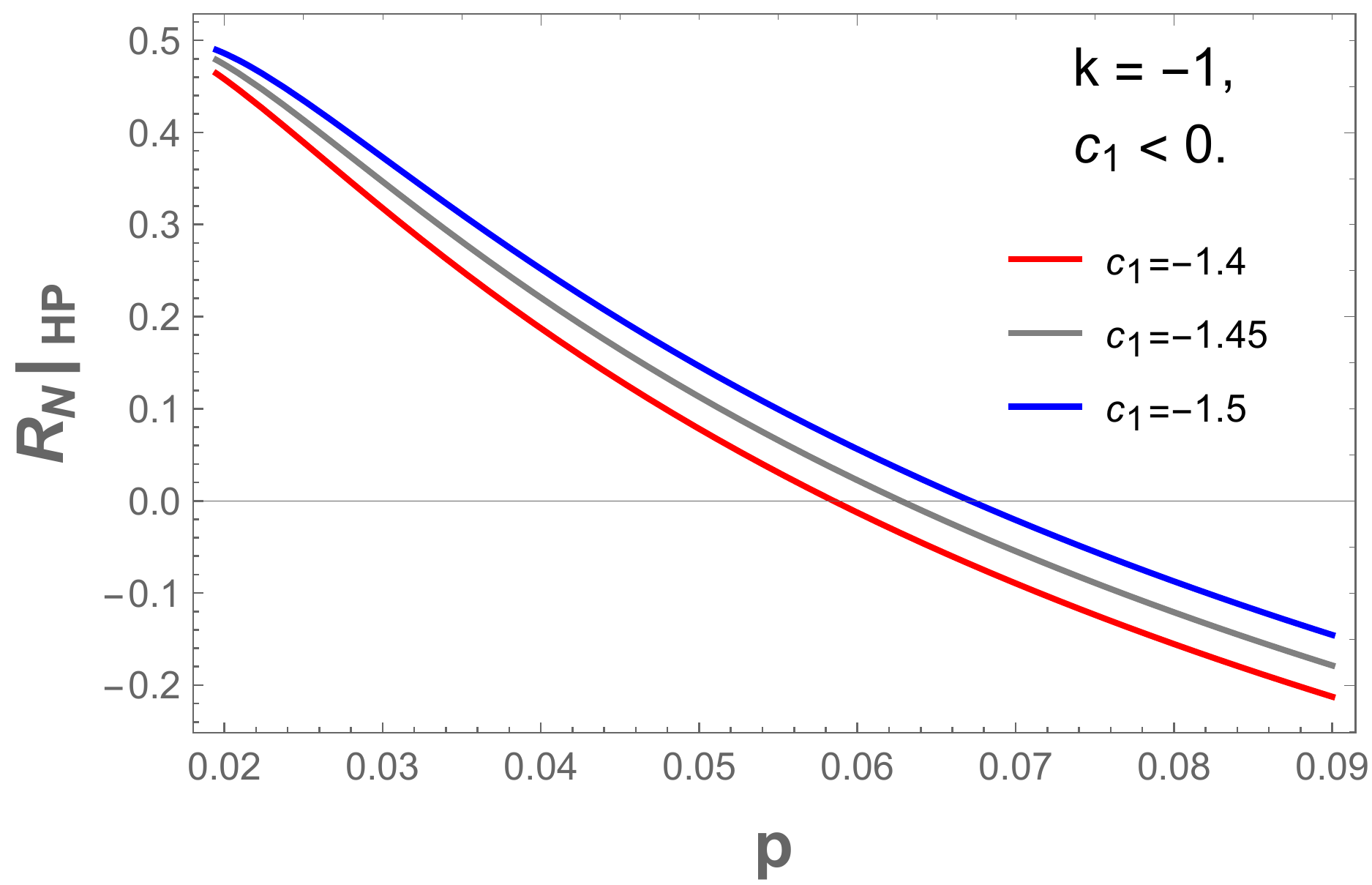}}
 		
 		\caption{\footnotesize In the case of massive coefficient $c_1 < 0$, for various horizon topologies $k$, the normalized scalar curvature at HP transition point (i.e.,  $R_N\big|_{HP}$)  as a function of pressure $p$  shows the sign changing behavior. (Here, the parameters $m =c_0=1, \ c_2=2$, are used.).  \label{fig:RNhp_c1<0}	}  
 	}  
 \end{figure}

\noindent
For $c_1 < 0$,  $R_N$ shows an interesting behavior, namely, it can be positive, negative and also vanish in the large black hole branch, i.e., there can respectively be repulsive, attractive and non-interacting situations, in both metastable and stable large black hole branches (See Fig.~\ref{fig:RN_c1<0}.).
The extremal black holes with $T=0$, have the normalized scalar curvature $R_N =\frac{1}{2}$, irrespective of horizon topology and graviton mass (see also~\cite{Yerra:2020tzg} for related discussion). Our next objective is to extract information about effective interactions of microstructures by evaluating  $R_N$ at the HP transition point.
This can be done by plugging the corresponding temperature and thermodynamic volume in the equation~\eqref{eq:RN}, which reads as
\begin{equation}\label{eq:RNHP}
R_N\big|_{HP} = -\Bigg(\frac{3}{2} + \frac{c_0 c_1 m^2}{4}\sqrt{\frac{6}{\pi \epsilon p}} + \frac{1}{8}\bigg(\frac{c_0 c_1 m^2}{4}\sqrt{\frac{6}{\pi \epsilon p}}\bigg)^2\Bigg).
\end{equation}
Setting $m=0$ in the above equation and taking $(k = +1)$ for spherical topology $R_N\big|_{HP} $ is a negative constant ($-3/2$), matching the claim in~\cite{Wei:2020kra} that for AdS Schwarzschild black holes thermodynamic curvature is a universal constant. However, in the present case in eqn. (\ref{eq:RNHP}), there is explicit dependence on  pressure $p$ and other parameters of the system even at HP transition point, and hence the universal constancy nature of thermodynamic curvature is broken by the graviton mass terms. 
Moreover,  the interactions at HP transition point heuristically suggested by $R_N\big|_{HP} $
are attractive for the massive coefficient $c_1 \geq 0$, and can change the sign if the massive coefficient is $c_1 < 0$ (See Fig.~\ref{fig:RNhp_c1<0}.). This again is in contrast to~\cite{Wei:2020kra}, where only the attractive type interactions are found at this point for spherical topology case.  
 \vskip 0.5 cm
 \noindent 
When the massive coefficient $c_1 < 0$, the point where the normalized scalar curvature vanishes is:
\begin{equation}
R_N\big|_{HP} = 0 \implies p =\frac{3c_0^2 c_1^2 m^4}{32\pi \epsilon} \equiv p^\ast.
\end{equation}
Thus, for $p >(<)  p^\ast $, the effective interactions at the HP transition point are attractive (repulsive). HP transition is like a threshold for formation of stable large black holes and since the nature of microstructures varies with massive gravity parameters, this process is expected to be quite non-trivial than the case of Schwarzschild black holes in Einstein gravity~\cite{Wei:2020kra}. 
 \begin{figure}[h!]
 	{\centering
 		\subfloat[]{\includegraphics[width=3in]{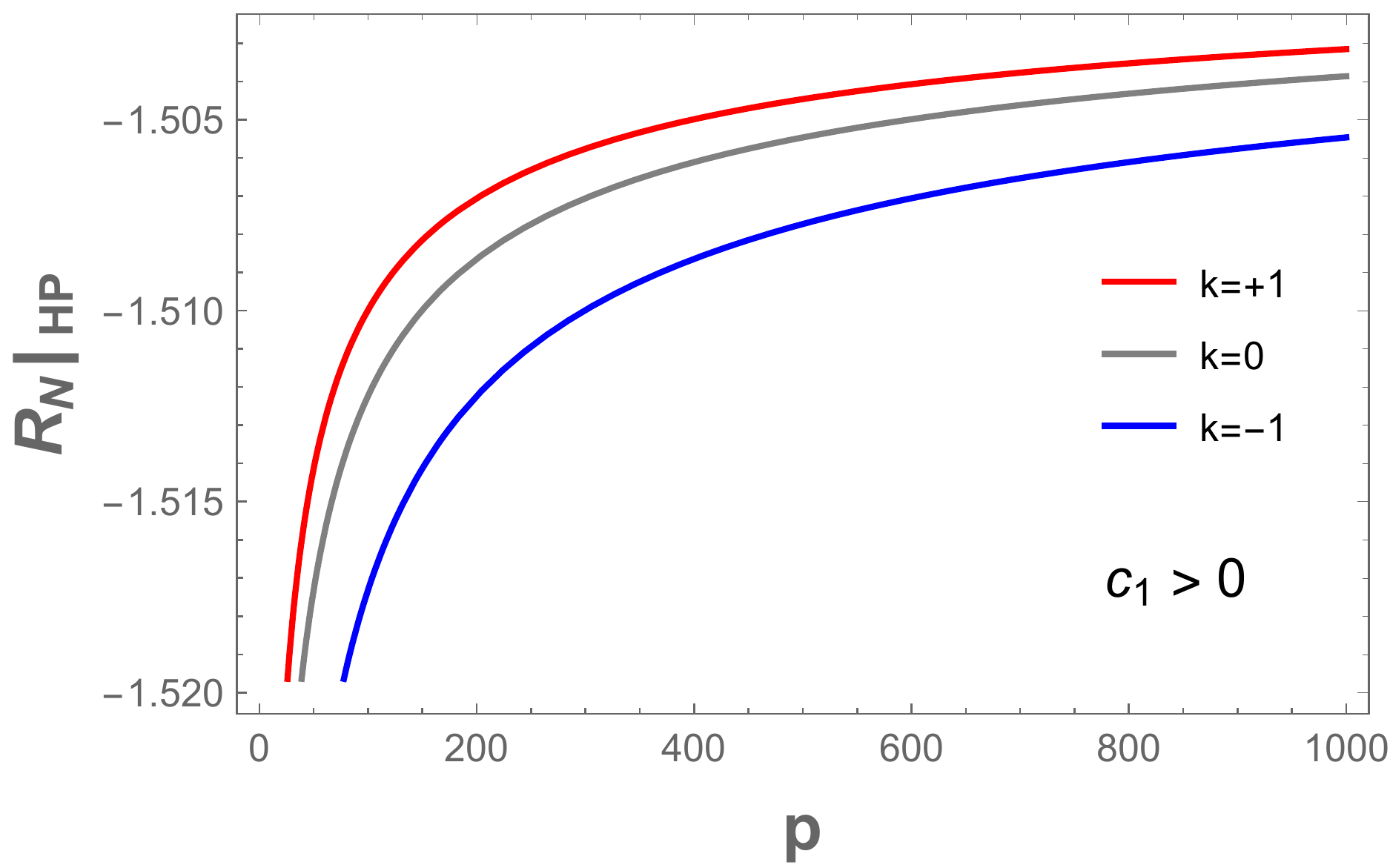}}\hspace{0.4cm}
 		\subfloat[]{\includegraphics[width=2.9in]{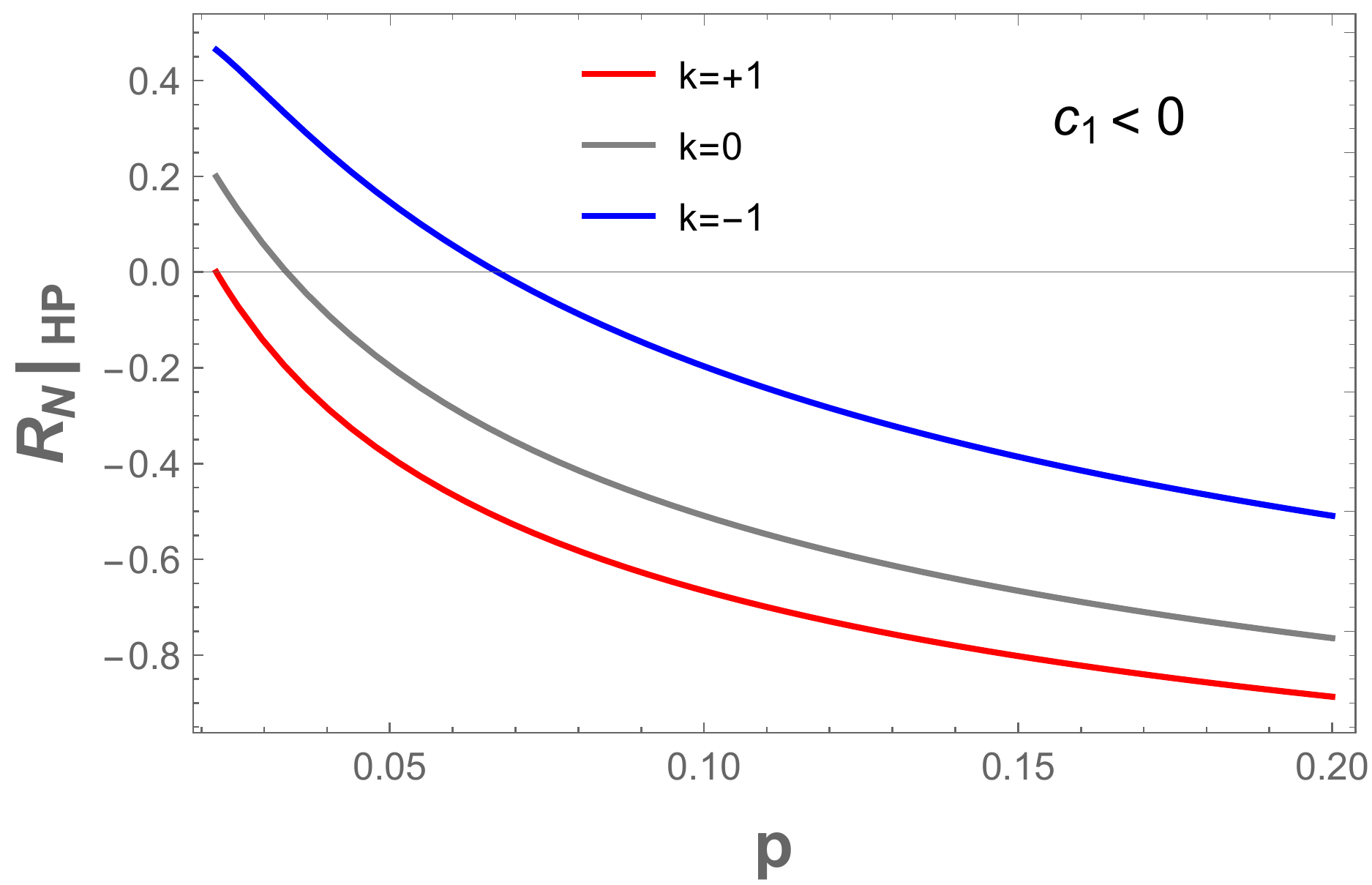}}
 		
 		\caption{\footnotesize The effect of horizon topology $k$ on the strength of  interactions at HP transition point (i.e., on  $\Big|R_N\big|_{HP}\Big|$), (a) for the massive coefficient $c_1 > 0$, and (b) for the massive coefficient $c_1 < 0$. (Here, the parameters $m =c_0=1, \ c_1=-1.5, \ c_2=2$, are used.).  \label{fig:RNhp_keff}	}  
 	}  
 \end{figure}  
Having learnt about the nature of interactions at the HP transition point, one can also comment on their strength, which can be inferred from the magnitude  $\Big|R_N|_{HP}\Big|$ for two different ranges of the parameter $c_1$. From equation-(\ref{eq:RNHP}), when the massive coefficient $c_1 > 0$, $\Big|R_N|_{HP}\Big|$ is highest for hyperbolic topology, followed by flat and spherical topology, i.e.,
\begin{equation}
\Big| R_N\big|_{HP} \Big|_{(k=-1)} > \Big| R_N\big|_{HP} \Big|_{(k=0)} > \Big| R_N\big|_{HP} \Big|_{ (k=+1)}.
\end{equation} 
However,  when the massive coefficient $c_1 < 0$ and for a certain range of parameters, the above order can be reversed, i.e., $\Big| R_N\big|_{HP} \Big|_{(k=-1)} < \Big| R_N\big|_{HP} \Big|_{(k=0)} < \Big| R_N\big|_{HP} \Big|_{(k=+1)}$. Some values of $c_1 <0$ and possible range of pressures where this happens are shown in Fig.~\ref{fig:RNhp_keff}. \\

\noindent
Towards the end of subsection-(\ref{phase}), possibility of HP transition at zero temperature was noted at a critical value of graviton mass $\hat{m}^2$.
The Ruppeiner scalar for the black holes with spherical and hyperbolic horizons at this zero temperature HP transition turns out to be  $R_N^{(T_{HP} = 0)} |_{HP}  = -3/2$ (from eqn. (\ref{eq:RNHP})), matching the value found in the massless graviton limit~\cite{Wei:2020kra}. This is a universal constant independent of horizon topology, other parameters and also graviton mass. Thus, the d-dimensional value of $R_N^{(T_{HP} = 0)} |_{HP}$ at $\hat{m}^2$ is also expected to be a universal negative constant (depending only on the space-time dimension), taking exactly the same value known for d-dimensional Schwarzschild black hole in AdS~\cite{Wei:2020kra}.

\section{Decoupled geometry at the HP transition point} \label{4}

We now take a closer look at the geometry near the HP transition point in Schwarzschild black holes in AdS and then generalize the results to corresponding solutions in massive gravity theory. 
Let us first compare the scaling of various quantities for charged and neutral black holes in AdS, at the second order critical point and HP transition point with respective parameters, as shown below.
\begin{center}
\begin{tabular}{c|c}
\hline	
Charged AdS black holes   & Schwarzschild-AdS black holes \\ 
at critical point & at HP transition point \\
	\hline \\
 $r_{cr} \sim  q^{1/(D-3) }$, &  $r_{\rm HP}^{\phantom{HP}} \sim  l $,  \\	\\
	$T_{cr} \sim q^{-1/(D-3)}$, & $T_{\rm HP}^{\phantom{HP}} \sim l^{-1}$,\\ \\	
	$p_{cr} \sim q^{-2/(D-3)}$. & $p_{\rm HP}^{\phantom{HP}} \sim l^{-2}$.\\  \\		
		\hline
\end{tabular} 
\end{center}
Here $D$ is the space-time dimension. Scaling of thermodynamic quantities at the HP transition point  \footnote{Schwarzschild-AdS black holes at minimum temperature $T_0$ also show the same scaling behavior as present at HP  transition point, which has in fact been used to propose a type of holographic duality in~\cite{Wei:2020kra} } with respect to AdS length $l$~\cite{Belhaj:2015hha} is similar to the scaling with respect to the charge $q^{1/(D-3)}$ at the critical point~\cite{Kubiznak:2012wp,Johnson:2017asf}.  For charged black holes in AdS, the proposal in~\cite{Johnson:2017asf} is that, as the critical point is approached in a large charge limit together with a near horizon limit, the black hole geometry decouples in to a d-dimensional Rindler space-time. Now, in the case of Schwarzschild black holes in AdS, an analogous double scaling limit at the HP transition point can now be thought of as the limit in which $l \rightarrow \infty$ taken together with the near horizon limit. To see this,
consider the metric of the d-dimensional Schwarzschild-AdS black hole geometry~\cite{Tangherlini:1963bw}:
\begin{equation}
ds^{2}=-Y(r) dt^{2}+\frac{dr^{2}}{Y(r) } + r^{2}\, d\Omega_{D-2}^2\, ,
  \label{eq:metricS}
\end{equation}
where the lapse function  is:
\begin{equation} \label{s}
	Y(r) =1-\frac{16 \pi M}{ (D-2)\omega_{D-2}} \frac{1}{r^{D-3}} + \frac{ r^{2}}{l^2} \ ,
	\end{equation}
and $\omega_{D-2}$ is the volume of the round $S^{D-2}$ surface. Naively, taking the $l \rightarrow \infty$ on the above black hole solution in eqn. (\ref{s}), will give us the corresponding solution in flat space-time, whose near horizon limit is the well known $ \text{Rindler}_{2} \times S^{D-2}$. Instead, by writing $r=r_++\eta\sigma$ and $t= \tau/\eta$ and for small $\eta$, the metric for the Schwarzschild-AdS black holes at the HP transition point becomes,
 \begin{equation}\label{eq:sch_metric_approx}
 ds^2 = -\Big(\frac{\sigma Y'(r_+)}{\eta}\Big)d\tau^2+ \frac{1}{\Big(\frac{\sigma Y'(r_+)}{\eta}\Big)} d\sigma^2+ (r_+^2 + 2\eta \sigma r_+)d\Omega_{D-2}^2\, ,
 \end{equation} 
  where, $Y(r=r_+)=0$ and $Y' = \frac{d Y(r)}{dr} |_{\rm(r=r_+)} = 4\pi T$. Now, at the HP transition point, we have $T = T_{\rm HP} = \frac{(D-2)}{2\pi l}$ and $r_+ = r_{\rm HP} =l$~~\cite{Belhaj:2015hha,Tangherlini:1963bw}. We can now take a new double scaling limit by approaching the horizon in the limit $\eta \rightarrow 0$, while at the same time taking the  limit $ l \rightarrow \infty$, by holding $ l\eta  = \tilde{l} $ fixed. Then the metric~\eqref{eq:sch_metric_approx} becomes:
   \begin{equation}\label{eq:sch_metric_HP}
   ds^2 = -{(4\pi {\widetilde T}_{\rm HP})\,\,\sigma}d\tau^2+\frac{1}{(4\pi {\widetilde T}_{\rm HP})}\frac{d\sigma^2}{\sigma }+d{\mathbb R}^{D-2}\, ,
   \end{equation}   
 where, ${\widetilde T}_{\rm HP} = T_{\rm HP}/ \eta$. Also,  in the limit $l \rightarrow \infty$, the cosmological constant $\Lambda =0$, and the metric on the round $S^{D-2}$ becomes flat (as the radius $r_{\rm HP}$ of $S^{D-2}$ diverges).  Thus, the metric~\eqref{eq:sch_metric_HP} represents the  line element for fully decoupled (as the throat length diverges) D-dimensional Rindler space-time with zero cosmological constant, exactly analogous to the one uncovered for the charged AdS black holes at critical point in~\cite{Johnson:2017asf}.  The Rindler space-time obtained here is also different from the standard $\text{Rindler}_{2} \times S^{D-2}$ that one gets as the near horizon limit of general non-extremal black holes.\\

\noindent
The extension of above results to the case of massive gravity in 4-dimensions~\cite{Cai2015,Hendi:2017fxp} can be implemented in an analogous manner, as the similarity in the scaling of thermodynamic quantities at the critical point with respect to charge, and, with respect to AdS length at the HP transition point, continues 
to hold, as seen below:
\begin{center}
	\begin{tabular}{c|c}
		\hline	
		Charged Topological AdS black holes   & Topological-AdS black holes in massive  \\ 
	in massive gravity	at critical point & gravity at HP transition point \\
		\hline \\
		$r_{cr} = \sqrt{\frac{6}{\epsilon}}q $, &  $r_{\rm HP}^{\phantom{HP}} = l \sqrt{\epsilon}  $,  \\	\\
		$T_{cr} = \frac{\epsilon^{\frac{3}{2}}}{ 3\sqrt{6} \pi q} + \frac{m^2c_0 c_1}{4\pi} $, & $T_{\rm HP}^{\phantom{HP}} = \frac{\sqrt{\epsilon}}{\pi l} +\frac{m^2 c_0 c_1}{4\pi} $,\\ \\	
		$p_{cr} = \frac{\epsilon^2}{96\pi q^2 } $. & $p_{\rm HP}^{\phantom{HP}} = \frac{3}{8\pi l^2} $.\\  \\		
		\hline
	\end{tabular} 	
\end{center}
Here, $\epsilon$ is defined in equation-(\ref{epsilon}).
To examine the near horizon geometry for topological AdS black hole in massive gravity  at  HP transition point, the metric in~\eqref{eq:metric} with corresponding values inserted becomes:
\begin{equation}\label{eq:metric_sch_massive_approx}
ds^2 = -\Big(\frac{4\pi \sigma T_{\rm HP}}{\eta} \Big) d\tau^2+\frac{1}{\Big(\frac{4\pi \sigma T_{\rm HP}}{\eta} \Big)} d\sigma^2+ (r_{\rm HP}^2 +2\eta \sigma r_{\rm HP})h_{ij} dx_i dx_j\, .
\end{equation} 
One can take the near horizon limit $\eta \rightarrow 0$  while at the same time keeping $l$ large, by holding  $l\eta = \tilde{l}$ fixed. However, in addition to the these limits, looking at the form of $T_{\rm HP}$ in equation (\ref{eq:Thp_massive}) one also needs to take the limit $c_1 \rightarrow 0$  to get a consistent near horizon metric, for the case of arbitrary topology\footnote{If instead, one takes the $m \rightarrow 0$ or $c_0 \rightarrow 0$, then, due to the requirement of $\epsilon > 0$ in eqn. (\ref{eq:Thp_massive}), HP transition point ceases to exist for arbitrary topology}.  Thus, the metric in (\ref{eq:metric_sch_massive_approx}) then goes over to:
\begin{equation}
ds^2 = -{(4\pi {\widetilde T}_{\rm HP})\,\,\sigma}d\tau^2+\frac{1}{(4\pi {\widetilde T}_{\rm HP})}\frac{d\sigma^2}{\sigma }+d{\mathbb R}^{2}\, ,
\end{equation} 
where, ${\widetilde T}_{\rm HP} = T_{\rm HP}/\eta$, and  both $l \eta = \tilde l $ and $ c_1/\eta = \tilde c_1$ are held fixed.  This new triple scaling limit results in a completely decoupled Rindler space-time with zero cosmological constant, matching the one obtained for charged black hole at critical point in~\cite{Yerra:2020bfx}. One also notes that a $l \rightarrow \infty$ limit cannot be taken in the case when the massive coefficient $c_1$ is negative, as the  temperature in this limit takes the value $T_{\rm HP}= \frac{c_0 \, c_1 m^2}{4 \pi }$, which is unphysical~\cite{Hendi:2017fxp}. Thus, when $c_1$ is negative, a non-trivial near horizon limit as discussed above, which gives a decoupled geometry as in~\cite{Johnson:2017asf}, does not in general exist.

\begin{figure}[h!]
	{\centering
		\subfloat[]{\includegraphics[width=2.9in]{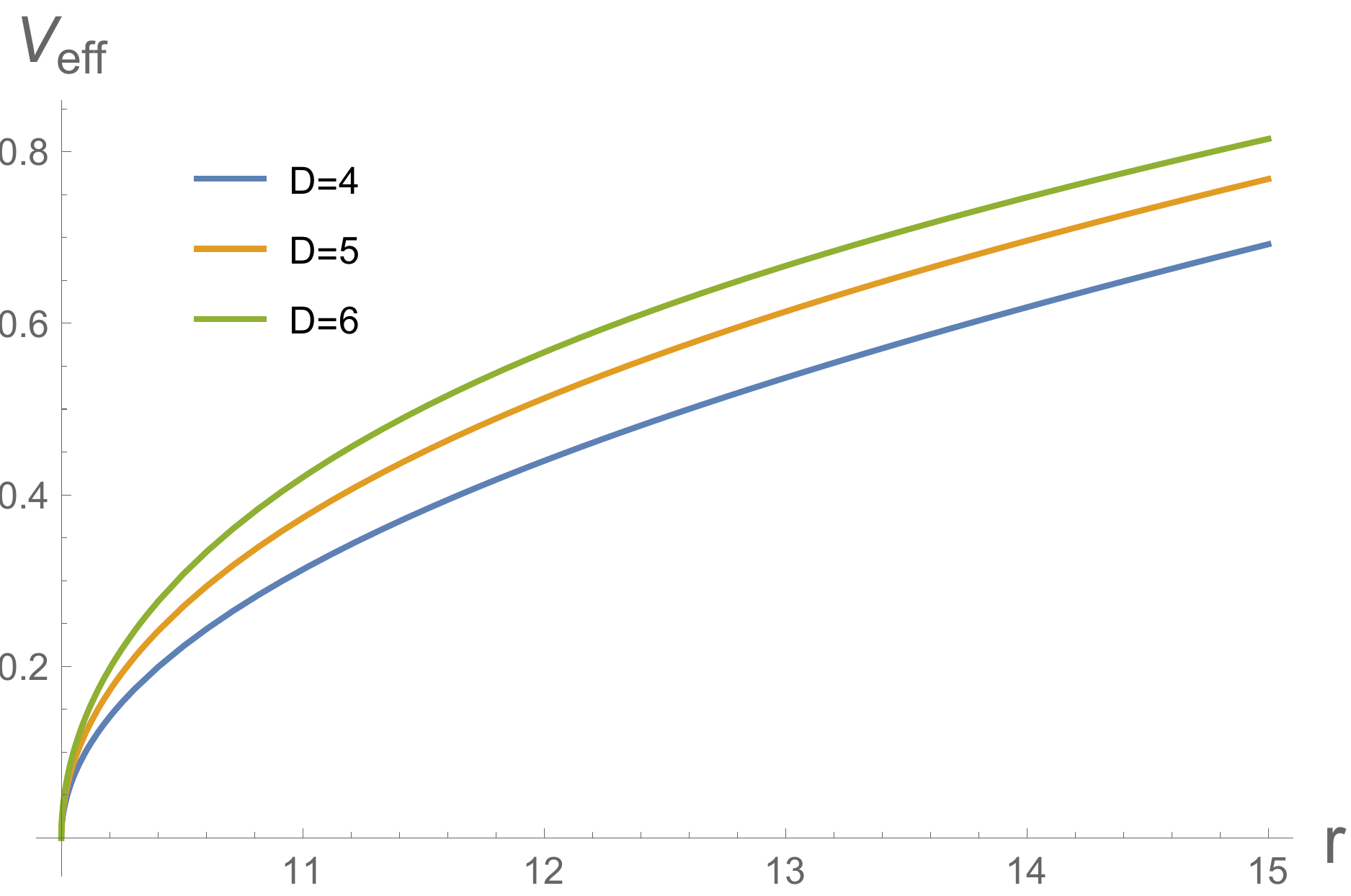} \label{fig:Veff_sch}}\hspace{0.4cm}
		\subfloat[]{\includegraphics[width=2.9in]{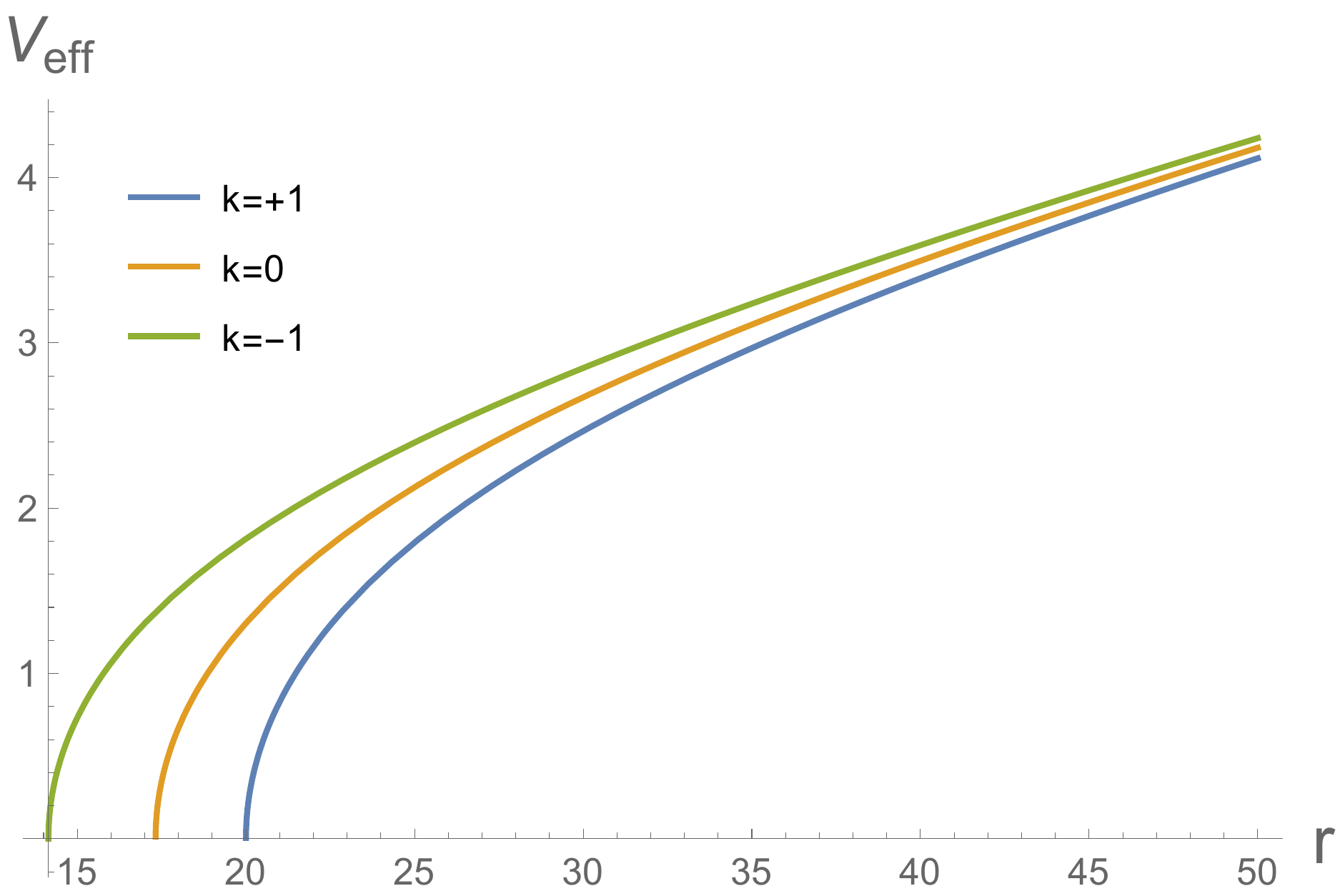}\label{fig:Veff_sch_massive}}
		
		\caption{\footnotesize The effective potential $V_{eff}$ with $L=0, \  l=10, $ and $\mu=0.5$, showing the absence of local minimum  (a) Schwarzschild-AdS black holes in various space-time dimensions $D$,  (b) 4-dimensional topological AdS black holes in massive gravity at various topologies $k$ (Here, the parameters $m=c_0 =1, \ c_1 =2, \ c_2 = 3,$ are used.).}  
	}  
\end{figure}

\vskip 0.5cm \noindent
There is one other way to inspect the geometry of black holes at the HP transition point, by studying the behavior of probe particles in its background. For instance, for the case of charged AdS black hole at the critical point, such a study was performed in \cite{Johnson:2017asf}, which gives information about the stability of the system.
Following the methods in~\cite{Johnson:2017asf,Chandrasekhar1984,doi:10.1142/S0217732311037261}, the effective potential for the motion of a point particle of mass $\mu$  moving in the background of the Schwarzschild-AdS black hole at the HP transition point is 
\begin{equation} \label{eq:Veff_sch}
V_{\rm eff}(r) = \sqrt{Y_{\rm HP}^{\phantom{HP}}(r)} \ \sqrt{\mu^2+\frac{L^2}{r^2}}\ ,
\end{equation}
where the lapse function is
\begin{equation} 
Y_{\rm HP}^{\phantom{HP}}(r) =1-\frac{16 \pi M_{\rm HP}}{ (D-2)\omega_{D-2}} \frac{1}{r^{D-3}} + \frac{ r^{2}}{l^2} \ ,
\end{equation}
and
\begin{equation}
M_{\rm HP} = \frac{(D-2) \omega_{D-2}}{8\pi}  l^{D-3},
\end{equation}
with $L$ denoting the angular momentum of the particle.  The effective potential in equation (\ref{eq:Veff_sch}), which may generally have a minimum at some value of $r_{min}$ ($> r_{\rm HP}$), depending on the values taken by $\mu$ and $L$. It was argued in~\cite{Johnson:2017asf}, that the presence of such a local minimum for a test particle at rest would lead to a condensation and possibly an instability of the black hole.  As can be seen from Fig.~\ref{fig:Veff_sch}, there is no local minimum and  the potential is purely attractive type binding all the subsystems that make up the black hole together.
A similar analysis can be repeated for topological AdS black hole in massive gravity at the HP transition point, considering the effective potential~\eqref{eq:Veff_sch} of the probe  with (using equation \eqref{Y(r)}),  
\begin{equation}
Y_{\rm HP}^{\phantom{HP}}(r) =k-\frac{2 M_{\rm HP}}{r}  + \frac{ r^{2}}{l^2} +m^2\big(\frac{c_0 c_1}{2}r+ c_0^2 c_2\big)\ ,
\end{equation}
and
\begin{equation}
M_{\rm HP} = l\epsilon \big(\sqrt{\epsilon} +\frac{m^2 c_0 c_1}{4} l \big).
\end{equation}
Fig.~\ref{fig:Veff_sch_massive} confirms that there is no instability.

\section{Conclusions}  \label{conclusions}

In this paper, we considered the 4-dimensional topological AdS black holes in massive gravity, 
where the presence of massive graviton  admits the Hawking-Page transition for the black holes with various horizon topologies denoted by the parameter $k$, provided $\epsilon \equiv (k+m^2c_2 c_0^2) > 0$~\cite{Cai2015}. Taking  the temperature T and thermodynamic volume V as the 
fluctuation variables, we computed the normalized Ruppeiner scalar curvature $R_N$ which is an important empirical indicator of nature and strength of interactions among the microstructures of a thermodynamic system, from a macroscopic view point.

\vskip 0.5cm \noindent 
$R_N$ diverges for the black hole with minimum temperature $T_0$. The small black hole branch is neglected as they are unstable, whereas for the large black hole branch, the nature of microstructures depends on the sign of $R_N$, which in turn is governed by the sign of massive coefficient $c_1$. In the case when $c_1 \geq 0$ and for the large black hole branch\footnote{Here, the behavior of $R_N$ w.r.t massive coefficient $c_1$ for large black hole branch, is same as that of the corresponding branch in charged black hole case~\cite{Yerra:2020oph}.},  $R_N$ is negative indicating the dominance of attractive type interactions among the black hole microstructures.  On the other hand when $c_1$ is negative,  $R_N$ can be positive, negative and also zero, from which one presumes the presence of repulsive, attractive, and vanishing interactions respectively, among the black hole microstructures. Interestingly, following up on a recent observation~\cite{Yerra:2020tzg}, we find that the microstructures of extremal black holes are quite unique, in the sense that the value of the Ruppeiner scalar is $R_N =\frac{1}{2}$, which is independent of their horizon topology and also graviton mass. 

\vskip 0.5cm \noindent $R_N$ evaluated at the Hawking-Page transition point gives an intriguing dependence on pressure, which is unexpected, considering the recent proposal that $R_N$ (computed for Schwarzschild-AdS black hole with spherical horizon topology $(k=+1)$ in massless graviton case) should probably be a universal constant at this point~\cite{Wei:2020kra}. Furthermore, even at the HP transition point $R_N$ negative in the case of massive coefficient $c_1 \geq 0 $, whereas in the case of $c_1 <0$, it can be negative, positive or zero. The zero of $R_N\big|_{HP}$ occurs at a critical pressure  $p=p^\ast \equiv \frac{3 c_0^2 c_1^2 m^4}{32\pi \epsilon}$ (or equivalently at a critical AdS length $l^\ast \equiv \frac{2 \sqrt{\epsilon}}{c_0 |c_1| m^2}$), such that the nature of interactions of microstructures are suggestively of attractive (repulsive) type for  $p >(<) \, p^\ast$. 
Since the nature of microstructures varies with parameters, the formation of black holes in massive gravity theories is much more non trivial than the simple explanation of repulsive interactions shifting to attractive type at HP transition in Schwarzschild black holes in Einstein gravity~\cite{Wei:2020kra}. 

\vskip 0.5cm \noindent  The strength of microstructure interactions at the HP transition (i.e., the magnitude  $\Big|R_N|_{HP}\Big|$) depends on the topology parameter $k$ and the massive gravity coefficient $c_1$. For $c_1 > 0$ one has,
\begin{equation*}
	\Big| R_N\big|_{HP} \Big|_{(k=-1)} > \Big| R_N\big|_{HP} \Big|_{(k=0)} > \Big| R_N\big|_{HP} \Big|_{ (k=+1)} \, .
\end{equation*} 
However, the order can be reversed, i.e., $\Big| R_N\big|_{HP} \Big|_{(k=-1)} < \Big| R_N\big|_{HP} \Big|_{(k=0)} < \Big| R_N\big|_{HP} \Big|_{(k=+1)}$, in the case of massive coefficient $c_1 <0$, for some range of pressures as shown in Fig.~\ref{fig:RNhp_keff}.

\vskip 0.5cm \noindent There are some special limits which are worth mentioning.  The universal constant nature of Ruppeiner scalar curvature $R_N$ at HP transition point\cite{Wei:2020kra} is broken due to the graviton mass and can be restored in a special limit of vanishing of the massive gravity coefficient $c_1$.  This is a case where one takes the parameters, $k=1, \ c_0=1, \ c_1=0,$ and $ c_2 =-1/2 $ and the system under consideration was called the \textit{holographic massive gravity model} of quantum field theories. Here, the breaking of diffeomorphism in bulk due to the generation of effective~\cite{Adams:2014vza}. In particular, there is a critical value of graviton mass where momentum dissipation effects are strong, leading to divergence of effective AdS length and also leading to a zero temperature confinement-deconfinement transition. We find that these phenomena persist in a more general case and also with hyperbolic topology, where HP transition can happen at zero temperature when the graviton mass takes a critical value $\hat{m}^2 = -k/(c_2 c_0^2)$. In this regard, $R_N^{(T_{HP} = 0)} |_{HP}$ at $\hat{m}^2$ is a universal negative constant (independent of graviton mass, see eqn.~\ref{eq:RNHP}), taking exactly the same value found earlier for d-dimensional Schwarzschild black hole in AdS~\cite{Wei:2020kra}. This intriguing result permits a speculation that the Ruppeiner scalar curvature $R_N$ along the zero temperature confinement-deconfinement transition in dual field theory should also be a constant and also be independent of the momentum-dissipation rate, indicating the strongly correlated nature of microstructures. It would be nice to directly compute $R_N$ in a holographic massive gravity model~\cite{Adams:2014vza,Alishahiha:2010bw} to verify this.

\vskip 0.5cm \noindent 
The geometry of the black hole close to the HP transition point was also explored in new double and triple near horizon scaling limits, respectively for the d-dimensional AdS Schwarzschild black hole and their counterparts in the massive gravity theory.  In both the cases, with appropriate parameter choices, the presence of a fully decoupled Rindler space-time was seen, which is different from the standard near horizon geometry of general non-extremal black holes. A study of probe particles moving in the background of black holes in AdS at the HP transition point shows that the system is stable against small perturbations.

\section*{Acknowledgements}
One of us (C.B.) thanks the DST (SERB), Government of India, for financial support through the Mathematical Research Impact Centric Support (MATRICS) grant no. MTR/2020/000135.

\bibliographystyle{apsrev4-1}
\bibliography{Novel@HP}
\end{document}